\documentclass[a4paper,cits]{JINST}
\usepackage{microtype}
\usepackage{booktabs}
\usepackage{amsmath}
\usepackage{lineno}

\let\ifpdf\relax

\newcommand{\bbonu}{\ensuremath{\beta\beta0\nu}}
\newcommand{\bbtnu}{\ensuremath{\beta\beta2\nu}}
\newcommand{\Qbb}{\ensuremath{Q_{\beta\beta}}}
\newcommand{\CS}{\ensuremath{^{137}}Cs}

\newcommand{\NA}{\ensuremath{^{22}}Na}

\newcommand{\XE}{\ensuremath{{}^{136}\rm Xe}}
\newcommand{\TL}{\ensuremath{{}^{208}\rm{Tl}}}
\newcommand{\BI}{\ensuremath{^{214}}Bi}

\author{
\mbox{The NEXT Collaboration}

V.~\'Alvarez,$^{a}$
F.I.G.~Borges,$^{b}$
S.~C\'arcel,$^{a}$
J.~Castel,$^{c}$
S.~Cebri\'an,$^{c}$
A.~Cervera,$^{a}$
C.A.N.~Conde,$^{b}$
T.~Dafni,$^{c}$
T.H.V.T.~Dias,$^{b}$
J.~D\'iaz,$^{a}$
M.~Egorov,$^{d}$
R.~Esteve,$^{e}$
P.~Evtoukhovitch,$^{f}$
L.M.P.~Fernandes,$^{b}$
P.~Ferrario,$^{a}$
A.L.~Ferreira,$^{g}$
E.D.C.~Freitas,$^{b}$
V.M.~Gehman,$^{d}$
A.~Gil,$^{a}$
A.~Goldschmidt,$^{d}$
H.~G\'omez,$^{c}$
J.J.~G\'omez-Cadenas,$^{a}$\thanks{Spokesperson (gomez@mail.cern.ch)}
D.~Gonz\'alez-D\'iaz,$^{c}$
R.M.~Guti\'errez,$^{h}$
J.~Hauptman,$^{i}$
J.A.~Hernando Morata,$^{j}$
D.C.~Herrera,$^{c}$
F.J.~Iguaz,$^{c}$
I.G.~Irastorza,$^{c}$
M.A.~Jinete,$^{h}$
L.~Labarga,$^{k}$
A.~Laing,$^{a}$\thanks{Co-corresponding author} ~
I.~Liubarsky,$^{a}$
J.A.M.~Lopes,$^{b}$
D.~Lorca,$^{a}$
M.~Losada,$^{h}$
G.~Luz\'on,$^{c}$
A.~Mar\'i,$^{e}$
J.~Mart\'in-Albo,$^{a}$
A.~Mart\'inez,$^{a}$
G.~Mart\'inez,$^{j}$
T.~Miller,$^{d}$
A.~Moiseenko,$^{f}$
F.~Monrabal,$^{a}$\thanks{Co-corresponding author} ~
C.M.B.~Monteiro,$^{b}$
F.J.~Mora,$^{e}$
L.M. Moutinho,$^{g}$
J.~Mu\~noz~Vidal,$^{a}$
H.~Natal da Luz,$^{b}$
G.~Navarro,$^{h}$
M.~Nebot-Guinot,$^{a}$
D.~Nygren,$^{d}$
C.A.B.~Oliveira,$^{d}$
R.~Palma,$^{l}$
J.~P\'erez,$^{k}$
J.L.~P\'erez~Aparicio,$^{l}$
J.~Renner,$^{d}$
L.~Ripoll,$^{m}$
A.~Rodr\'iguez,$^{c}$
J.~Rodr\'iguez,$^{a}$
F.P.~Santos,$^{b}$
J.M.F.~dos~Santos,$^{b}$
L.~Segu\'i,$^{c}$
L.~Serra,$^{a}$
D.~Shuman,$^{d}$
A.~Sim\'on,$^{a}$
C.~Sofka,$^{n}$
M.~Sorel,$^{a}$
J.F.~Toledo,$^{d}$
A.~Tom\'as,$^{c}$
J.~Torrent,$^{m}$
Z.~Tsamalaidze,$^{f}$
J.F.C.A.~Veloso,$^{g}$
J.A.~Villar,$^{c}$
R.~Webb,$^{n}$
J.T.~White,$^{n}$
N.~Yahlali$^{a}$\\
\llap{$^{a}$}
Instituto de F\'isica Corpuscular (IFIC), CSIC \& Universitat de Val\`encia\\
Calle Catedr\'atico Jos\'e Beltr\'an, 2, 46980 Paterna, Valencia, Spain\\
\llap{$^{b}$}
Departamento de Fisica, Universidade de Coimbra\\
Rua Larga, 3004-516 Coimbra, Portugal\\
\llap{$^c$}
Lab.\ de F\'isica Nuclear y Astropart\'iculas, Universidad de Zaragoza\\ 
Calle Pedro Cerbuna, 12, 50009 Zaragoza, Spain\\
\llap{$^d$}
Lawrence Berkeley National Laboratory (LBNL)\\
1 Cyclotron Road, Berkeley, California 94720, USA\\
\llap{$^{e}$}
Instituto de Instrumentaci\'on para Imagen Molecular (I3M), Universitat Polit\`ecnica de Val\`encia\\ 
Camino de Vera, s/n, Edificio 8B, 46022 Valencia, Spain\\
\llap{$^{f}$}
Joint Institute for Nuclear Research (JINR)\\
Joliot-Curie 6, 141980 Dubna, Russia\\
\llap{$^{g}$}Institute of Nanostructures, Nanomodelling and Nanofabrication (i3N), Universidade de Aveiro\\
Campus de Santiago, 3810-193 Aveiro, Portugal\\
\llap{$^{h}$}
Centro de Investigaciones, Universidad Antonio Nari\~no\\ 
Carretera 3 este No.\ 47A-15, Bogot\'a, Colombia\\
\llap{$^{i}$}
Department of Physics and Astronomy, Iowa State University\\
12 Physics Hall, Ames, Iowa 50011-3160, USA\\
\llap{$^{j}$}
Instituto Gallego de F\'isica de Altas Energ\'ias (IGFAE), Univ.\ de Santiago de Compostela\\
Campus sur, R\'ua Xos\'e Mar\'ia Su\'arez N\'u\~nez, s/n, 15782 Santiago de Compostela, Spain\\
\llap{$^{k}$}
Departamento de F\'isica Te\'orica, Universidad Aut\'onoma de Madrid\\
Ciudad Universitaria de Cantoblanco, 28049 Madrid, Spain\\
\llap{$^{l}$}
Dpto.\ de Mec\'anica de Medios Continuos y Teor\'ia de Estructuras, Univ.\ Polit\`ecnica de Val\`encia\\
Camino de Vera, s/n, 46071 Valencia, Spain\\
\llap{$^{m}$}
Escola Polit\`ecnica Superior, Universitat de Girona\\
Av.~Montilivi, s/n, 17071 Girona, Spain\\
\llap{$^{n}$}
Department of Physics and Astronomy, Texas A\&M University\\
College Station, Texas 77843-4242, USA\\

E-mail: \email{andrew.laing@ific.uv.es, francesc.monrabal@ific.uv.es}
}
\title{Operation and first results of the NEXT-DEMO prototype using a
  silicon photomultiplier tracking array}

\abstract{NEXT-DEMO is a high-pressure xenon gas TPC which acts as a technological
  test-bed and demonstrator for the NEXT-100 neutrinoless double beta
  decay experiment. In its current configuration the apparatus fully implements
  the NEXT-100 design concept. This is 
  an asymmetric TPC, with an energy plane made of
  photomultipliers and a tracking plane made of silicon photomultipliers (SiPM) coated with TPB. The
  detector in this new configuration has been used to reconstruct the characteristic signature of
  electrons in dense gas. Demonstrating the ability to identify the
  MIP and ``blob'' regions. Moreover, the SiPM tracking plane allows
  for the definition of a large fiducial region in which an excellent
  energy resolution of 1.82\% FWHM at 511~keV has been measured (a
  value which extrapolates to 0.83\% at the xenon \Qbb).}

\preprint{arXiv:1306.0471}

\begin{document}

%\tableofcontents

%%% SECTION 1. INTRODUCTION
\section{Introduction} 
\label{sec:Introduction}
The NEXT-100 \cite{Alvarez:2012haa} time projection chamber (TPC),
currently under construction at \emph{Laboratorio Subterr\'aneo de
  Canfranc} (LSC), will search for neutrinoless double beta
decay (\bbonu) using 100--150~kg of xenon gas enriched in the \XE\
isotope to $91\%$. The detector boasts two important features for
\bbonu\ searches: \emph{excellent energy resolution} (better than 1\%
FWHM at the $Q$ value of \XE) and event \emph{topological information}
for the identification of signal and background. This combination
gives NEXT an excellent experimental sensitivity to \bbonu\ searches \cite{GomezCadenas:2011it}. In addition, the technique can be
extrapolated to the ton-scale, thus allowing the full exploration of
the inverted hierarchy of neutrino masses \cite{GomezCadenas:2012jv}.
%%%%%%%%%%
\begin{figure}
\centering
\includegraphics[width=0.8\textwidth]{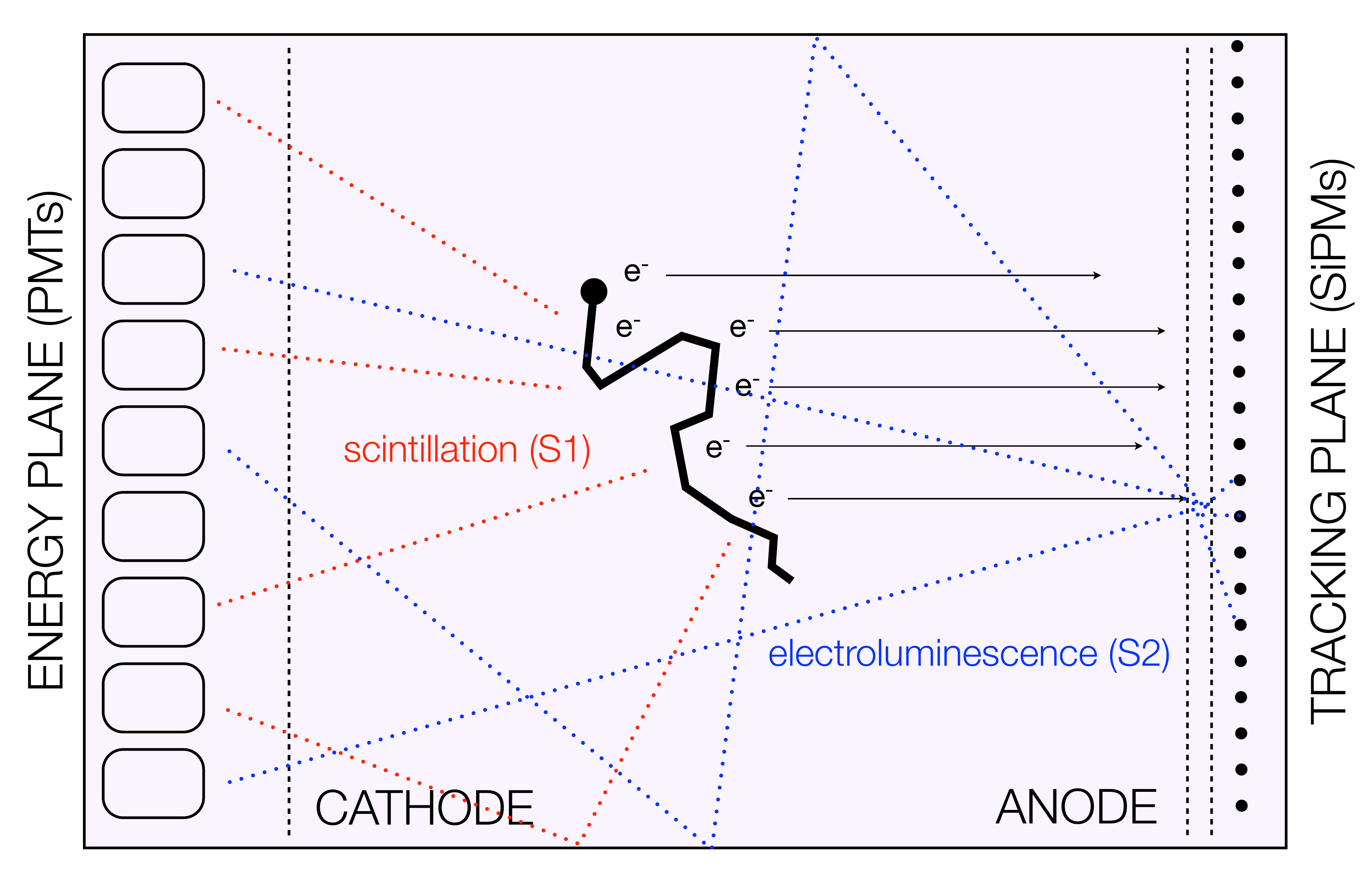}
\caption{The design of NEXT-100, implementing the \emph{Separate, Optimized Functions} (SOFT) concept: EL light generated at the anode is recorded in the photosensor plane right behind it and used for tracking; it is also recorded in the photosensor plane behind the transparent cathode and used for a precise energy measurement.} \label{fig:SOFT}
\end{figure}
%%%%%%%%%%

Energy resolution in NEXT is optimized by taking advantage of the electroluminescence (EL) of xenon. Using this effect, linear amplification of the ionization is possible allowing energy resolution to approach
that predicted by the low Fano factor of xenon gas (0.3\% FWHM at \Qbb). Furthermore, following ideas introduced in \cite{Nygren:2009zz} and further developed in \cite{Alvarez:2011my}, the chamber will have separate detection systems for tracking and calorimetry. NEXT-100 has chosen a simple and robust implementation of this \emph{SOFT} concept (\emph{Separate, Optimized Functions}), that of an asymmetric TPC with an energy plane made of photomultipliers (PMTs) behind a transparent cathode, and a tracking plane made of silicon photomultipliers (SiPM) behind the EL grids and anode. The design is illustrated in figure~\ref{fig:SOFT}. 

The detection process is as follows: Particles interacting in the HPXe transfer their energy to the medium through ionization and excitation. The excitation energy is seen as the prompt emission of VUV ($\sim$178~nm) scintillation light. The electron-ion pairs created by the passage of the particle are prevented from recombining by an electric field (0.3--0.5 ${\rm kV}~{\rm cm}^{-1}$) which causes the drift of the electrons towards the TPC anode. Before reaching the anode, the electrons pass through a mesh (76\% transparent) which defines the start of a region of more intense electric field (2~${\rm kV}~{\rm cm}^{-1}~{\rm bar}^{-1}$) which extends over the remaining 0.5~cm of drift to the 88\% open anode which is held at ground. In this region the electrons are induced to excite the gas producing further VUV photons via the process of electroluminescence (EL). The start-of-event is defined by the detection of the primary scintillation light in the energy plane with EL light being detected both in the energy plane where the ionization energy is accurately measured and in the tracking plane where the topology of the event can be reconstructed by the SiPMs. While both planes can operate essentially independently, both measurements are improved by considering the timing sensitivity of the PMTs and the position sensitivity of the SiPMs as is explained in sections \ref{subsec:evtTop} and \ref{sec:engRes}.

NEXT-DEMO was designed as a proof of concept and test-bed for the technology of the NEXT-100 detector. Its design is described in section \ref{sec:DEMO} with a more detailed description as well as initial energy resolution measurements using two PMT planes being described in \cite{Alvarez:2012xda,Alvarez:2012hu} .
Its current configuration utilizes a new tracking plane
comprised of 256 1-mm$^2$~SiPMs mounted with a pitch of 1~cm and coated
with \emph{tetraphenyl butadiene} (TPB). This detector geometry represents a scale model of the
current design for NEXT-100 and, as such, the results and methods
described here can be used as a direct indication of the performance
of the ultimate experiment and to finalize the ultimate design.

\section{NEXT-DEMO}
\label{sec:DEMO}
\begin{figure}
\centering
\includegraphics[width=0.75\textwidth,height=8cm]{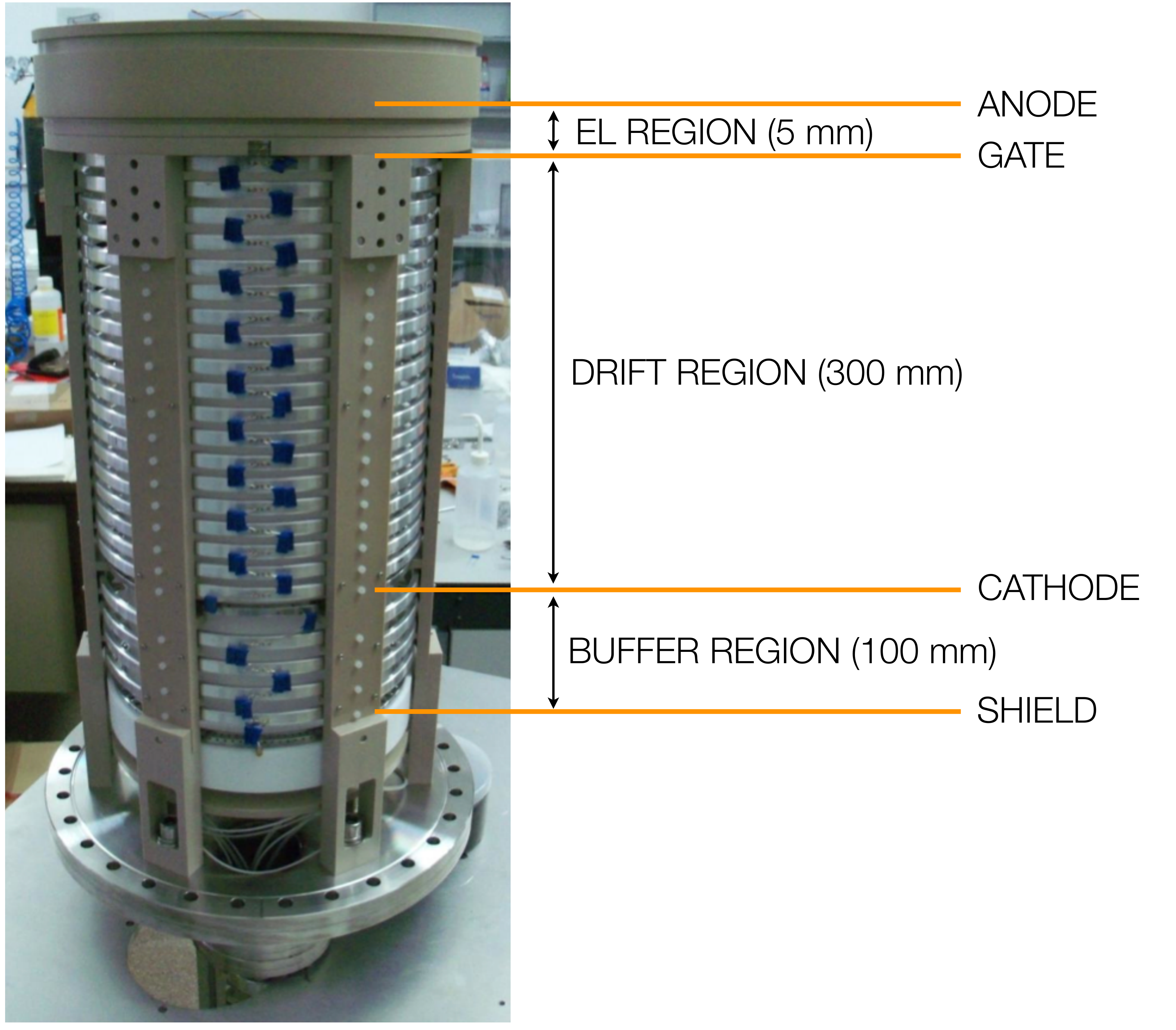}
\caption{External view of the time projection chamber standing on one end-cap. The approximate positions of the different regions of the TPC are indicated.} \label{fig:TPC}
\end{figure}

%%%%%%%%%%%%%%%%%%%%%%%%%%%%%%%%%%%%%%%%%%%%%%%%%%%%%%%%%%%%
NEXT-DEMO is a high-pressure xenon TPC contained within a cylindrical stainless-steel pressure vessel of diameter 30~cm and length 60~cm which was designed to withstand up to 15~bar (the data decribed used 10~bar). The TPC itself is defined by three metallic wire grids --- called \emph{cathode}, \emph{gate} and \emph{anode} --- which define the two active regions: the 30-cm long \emph{drift region}, between cathode and gate with a typical drift field of 500~V~cm$^{-1}$; and the 0.5-cm long \emph{EL region}, between gate and anode.  The electric field is created by supplying a large negative voltage to the cathode, then degrading it using a series of metallic rings of 30 cm diameter spaced 5 mm and connected via 0.5~G$\Omega$ resistors (shown in figure~\ref{fig:TPC}).  The gate is at negative voltage so that a moderate electric field --- for the data described here 2.0~$\mathrm{kV~cm^{-1}~bar^{-1}}$ --- is created between the gate and the anode, which is at ground. A set of six panels made of PTFE (Teflon) coated with TPB are mounted inside the electric-field cage forming a \emph{light tube} of hexagonal cross section with and apothem length of 8~cm.

%%%%%%%%%%
%%%%%%%%%%

%%%%%%%%%%
\begin{figure}
\centering
\includegraphics[scale=.7]{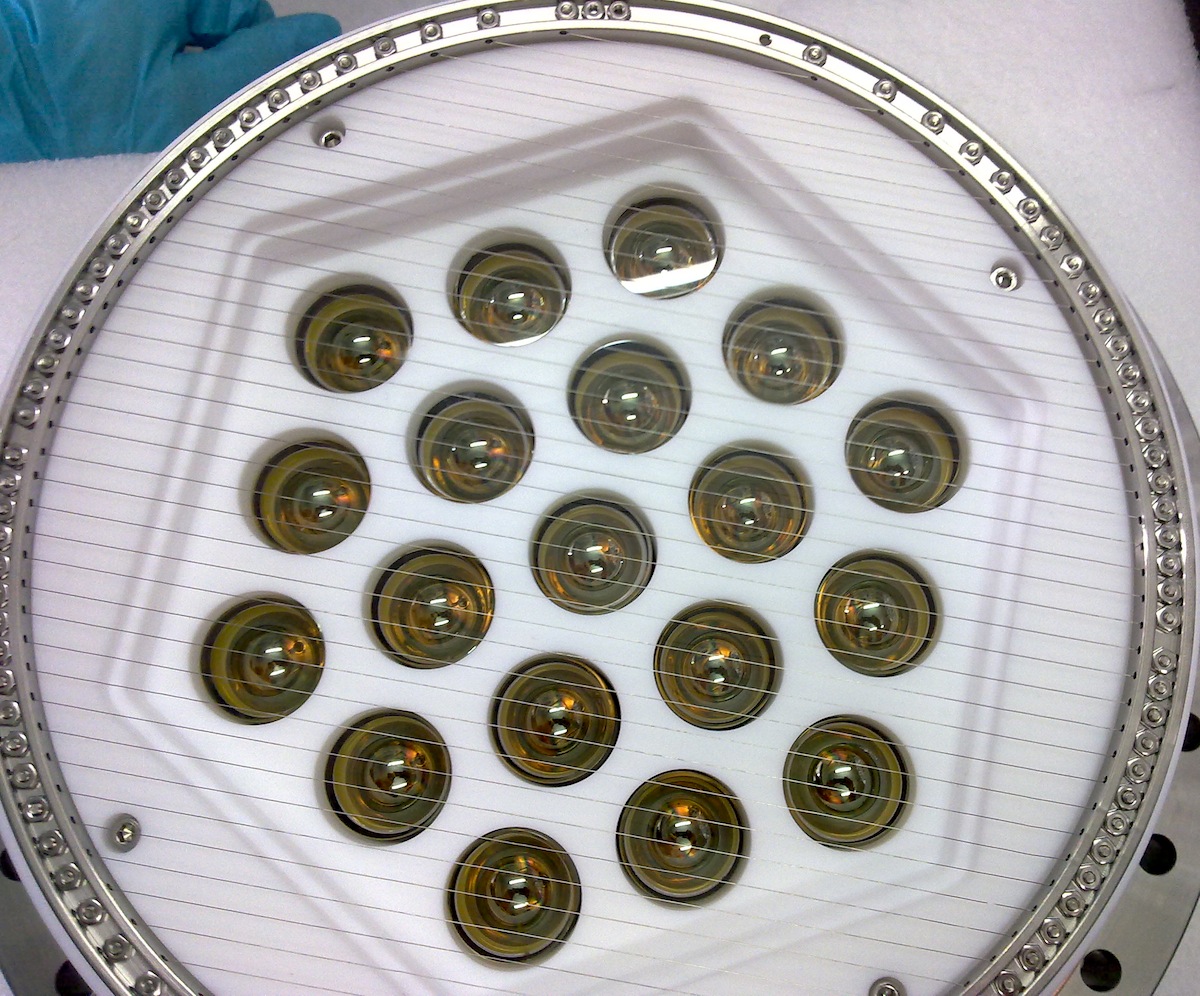}
\caption{The energy plane of NEXT-DEMO, equipped with 19 Hamamatsu R7378A PMTs.} \label{fig:DetPlanes}
\end{figure}
%%%%%%%%%%

The energy plane (see figure~\ref{fig:DetPlanes}) is equipped with 19 Hamamatsu R7378A photomultiplier tubes. These are 1-inch, pressure-resistant (up to 20 bar) PMTs with acceptable quantum efficiency ($\sim 15$\%) in the VUV region and higher efficiency at TPB wavelengths ($\sim 25$\%). The resulting photocathode coverage of the energy plane is about 39\%. The PMTs are inserted into a PTFE holder following a hexagonal pattern. A grid, known as \emph{shield} and similar to the cathode but with the wires spaced 0.5~cm apart, is screwed on top of the holder and set to $+500$~V. This protects the PMTs from the high-voltage of the cathode, and ensures that the electric field in the 10-cm buffer region is below the EL threshold. A more detailed description of the detector is available in~\cite{Alvarez:2012xda}. The tracking plane is described in more detail in the following section.

\section{The NEXT-DEMO tracking plane} 
\label{sec:tracking}
\subsection{Electron topology in HPXe gas}
\label{subsec:TOPO}
A major advantage of the NEXT design is the possibility of reconstructing the topology of an event in the detector. A \bbonu\ signal event involves 2 electrons, the energies of which sum to \Qbb\ ($\sim$2.45~MeV). While good energy resolution is enough to separate such events from the more abundant \bbtnu\ electrons, backgrounds still exist in the form of the compton and photoelectric interactions of high-energy photons emitted after the decay of \TL\ and \BI\ which result in single electrons with energy similar to \Qbb.
%%%%%%%%%%
\begin{figure}
  \begin{center}
    $\begin{array}{cc}
      \includegraphics[width=0.495\textwidth]{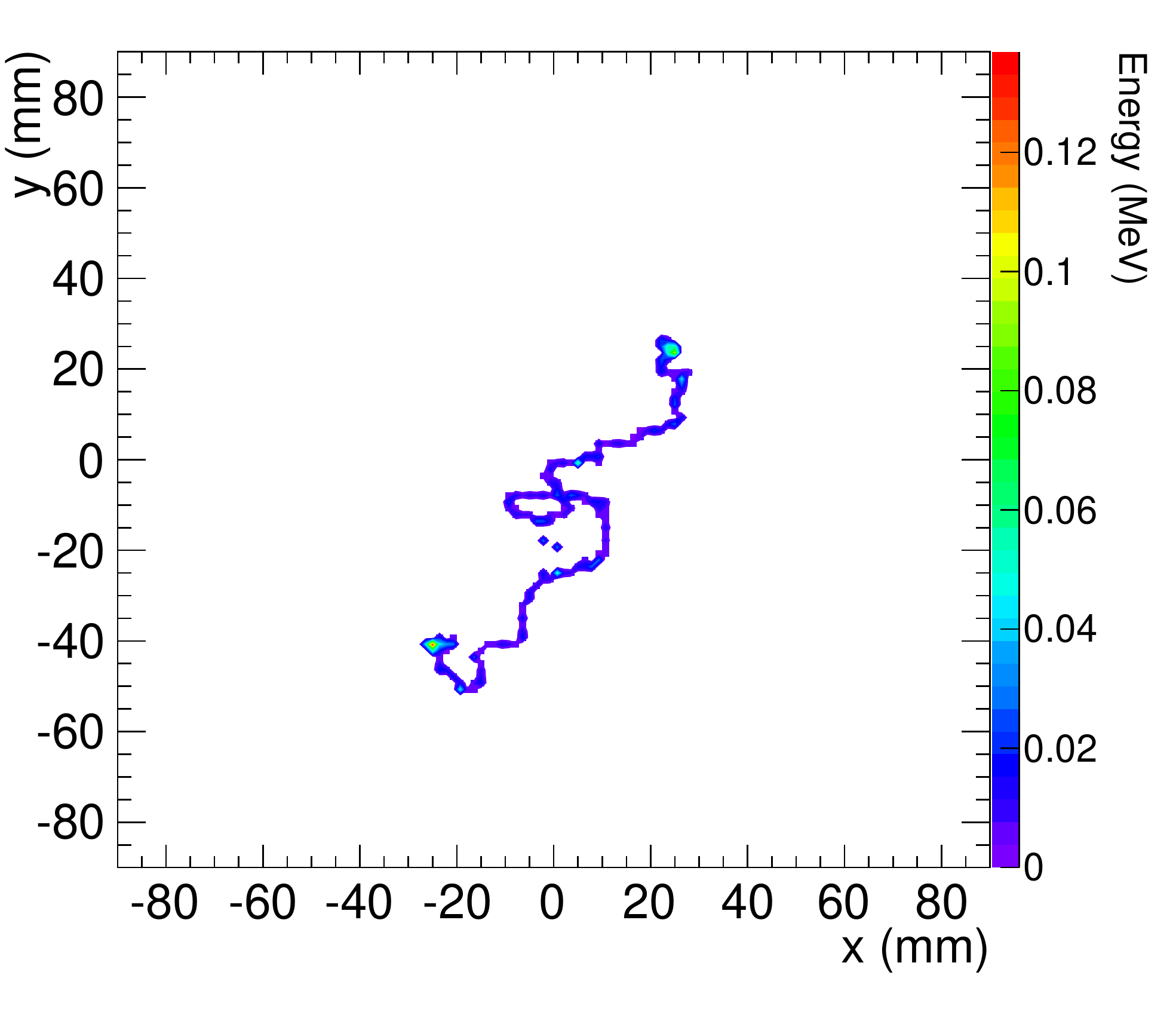} &
      \includegraphics[width=0.495\textwidth]{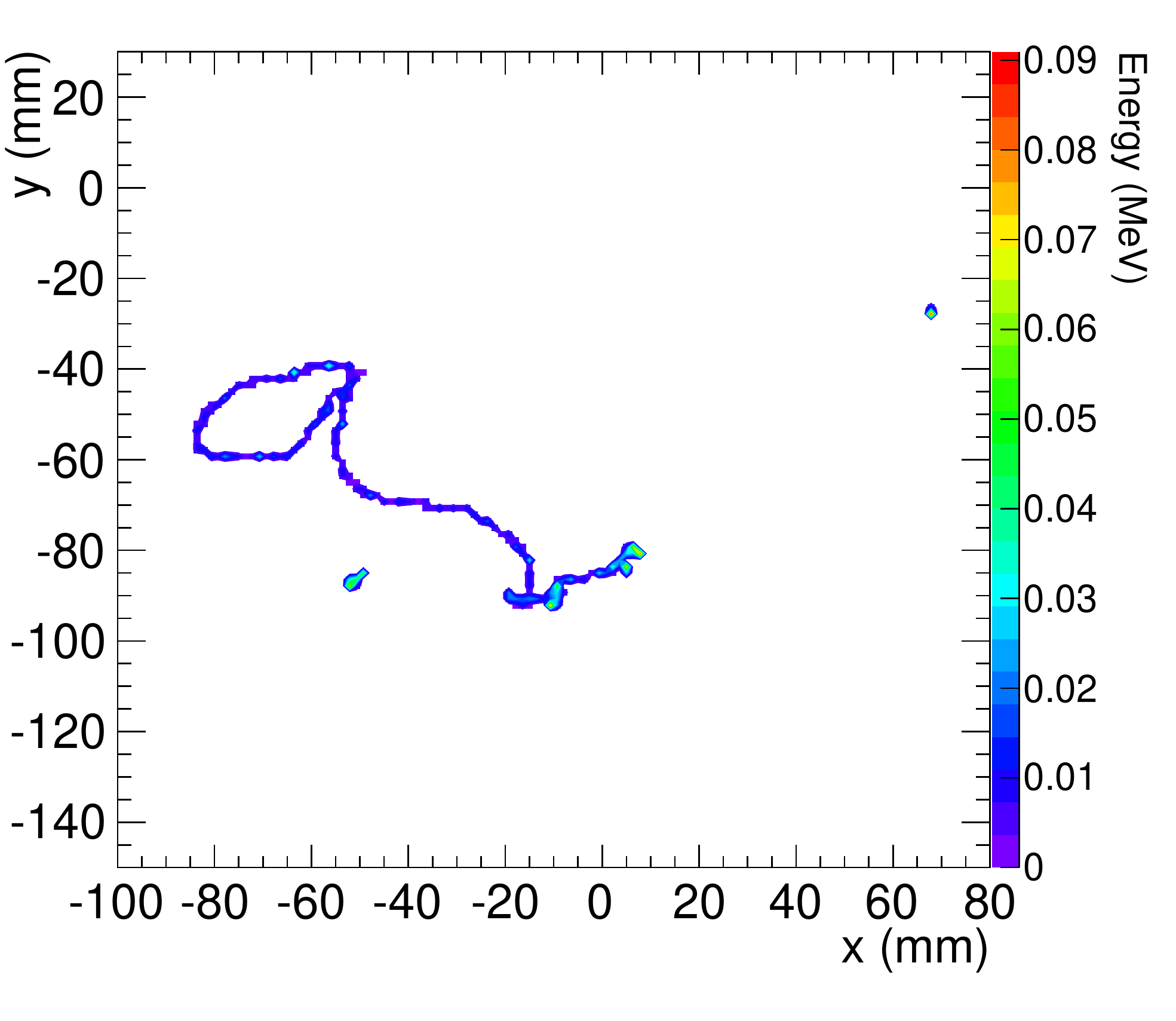}
    \end{array}$
    \caption{Topological signatures left by interactions in NEXT. Left: Monte-Carlo simulation of a \XE\ \bbonu\ event in xenon gas at 15 bar. Right: Monte-Carlo simulation of a background event, in this case an electron produced via Compton scattering of a 2.447~MeV photon emitted by \BI.}
    \label{fig:track}
  \end{center}
\end{figure}
%%%%%%%%%%
An electron passing through xenon gas at 10~bar will lose energy, generally, in two distinct phases: a MIP-like region with $dE/dx \simeq 70~{\rm keV}~{\rm cm}^{-1}$ and, a region at the end of the track dominated by multiple scattering where $\sim$300~keV is deposited in a short distance (a ``blob''). In order to separate a signal event from one induced by the backgrounds described above, a topology of two electrons with a common vertex must be distinguished from a single electron track. This requires the power to resolve the ``blob'' and MIP regions as well as separate an end-point blob energy deposit from a fluctuation in the $dE/dx$ of the MIP region induced, for example, by bremsstrahlung or delta rays emitted by the electrons. Therefore, in designing the tracking plane a number of physical effects, both inherent to the interaction of electrons in the gas and resulting from the read-out of the signal, must be taken into account. The most important of these are the emission of delta electrons and bremsstrahlung photons, and the diffusion of the electron cloud during drift.

The tracks shown in figure~\ref{fig:track} are those of a simulated signal event and a compton electron liberated by a 2.447~MeV gamma from \BI\ where the energy scale shown is that of the energy deposited by the electron and the effects of drift are not included. In both cases the end-point ``blob'' regions are clearly visible as high energy deposits at the extremes of the tracks. However, both exhibit additional deposits of energy greater than that predicted by MIP energy loss within the track as well as deposits which are separated from the main trace of the electrons. While sufficient position resolution is required to separate the ``blob'' and MIP regions and so that regions of increased $dE/dx$ are not added together to the extent that they can mimic the deposit of an additional ``blob'', very fine grained tracking is of little use since the blurring of the track due to photon and delta electron emission and the subsequent drift of the charge cloud remove any useful information at this level. 
In pure xenon, the transverse diffusion expected is of order 1~mm~cm$^{-\frac{1}{2}}$ and the read-out plane must be positioned a few mm from the EL production region which results in blurring of order 1~cm at read-out. Taking all these effects into account, SiPM are a good technological choice since they allow for the use of moderate EL without problems of stability, they have a relatively low cost and, a reasonable position resolution can be achieved without very dense arrays. Detailed simulation studies \cite{Alvarez:2011my} yielded a pitch of 1~cm as the best compromise to provide a good separation between the two blobs of the signal electron and an acceptable number of channels at 10-15~bar of pure xenon. The tracking plane of NEXT-DEMO was built, consequently, with that pitch. 

\subsection{Tracking plane design}
\label{subsec:trackDes}
\begin{figure}
  \begin{center}
  \includegraphics[width=0.475\textwidth,height=5cm]{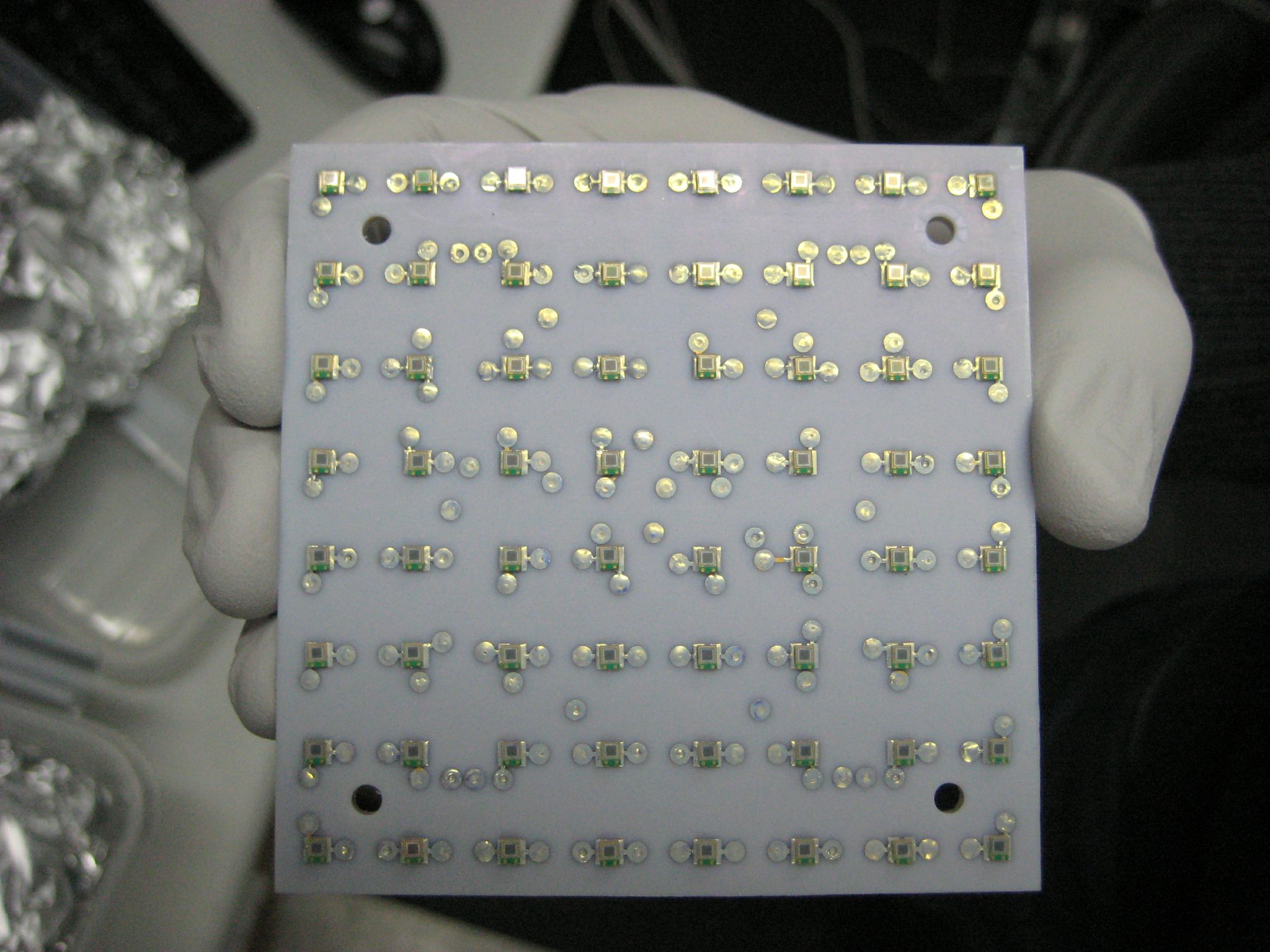}
  \includegraphics[width=0.475\textwidth,height=5cm]{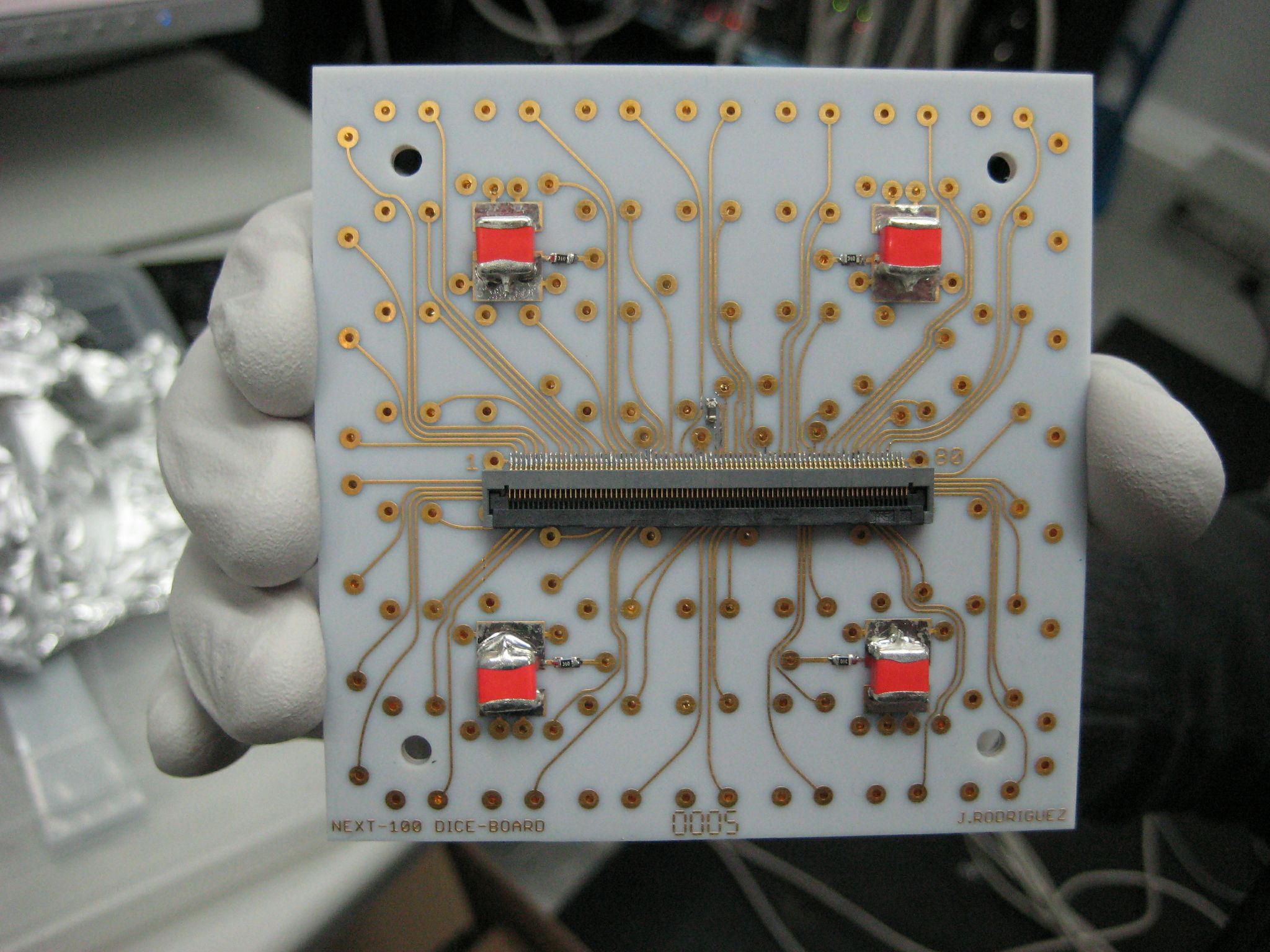}
  \includegraphics[width=0.475\textwidth,height=5cm]{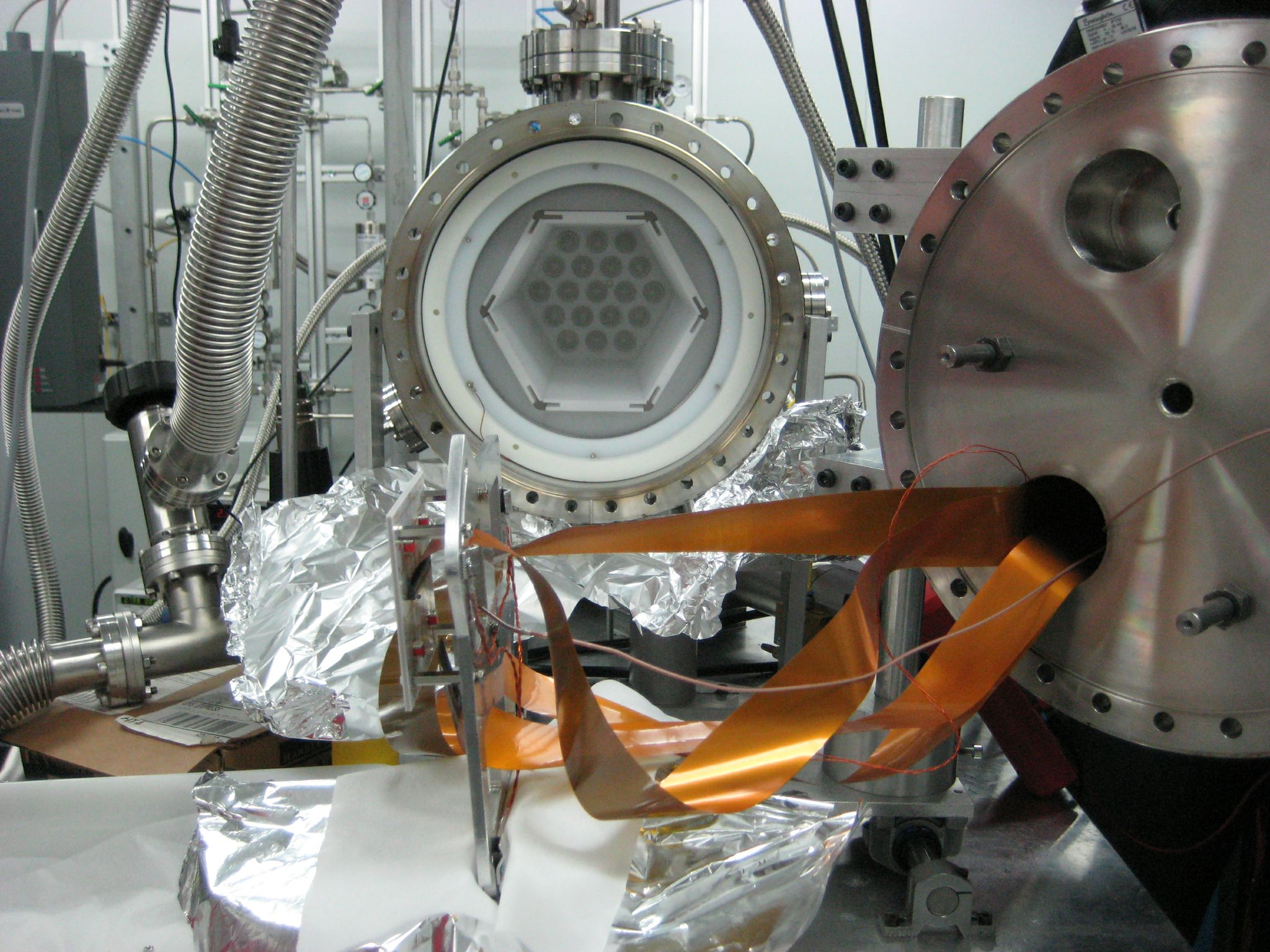}
  \includegraphics[width=0.475\textwidth,height=5cm]{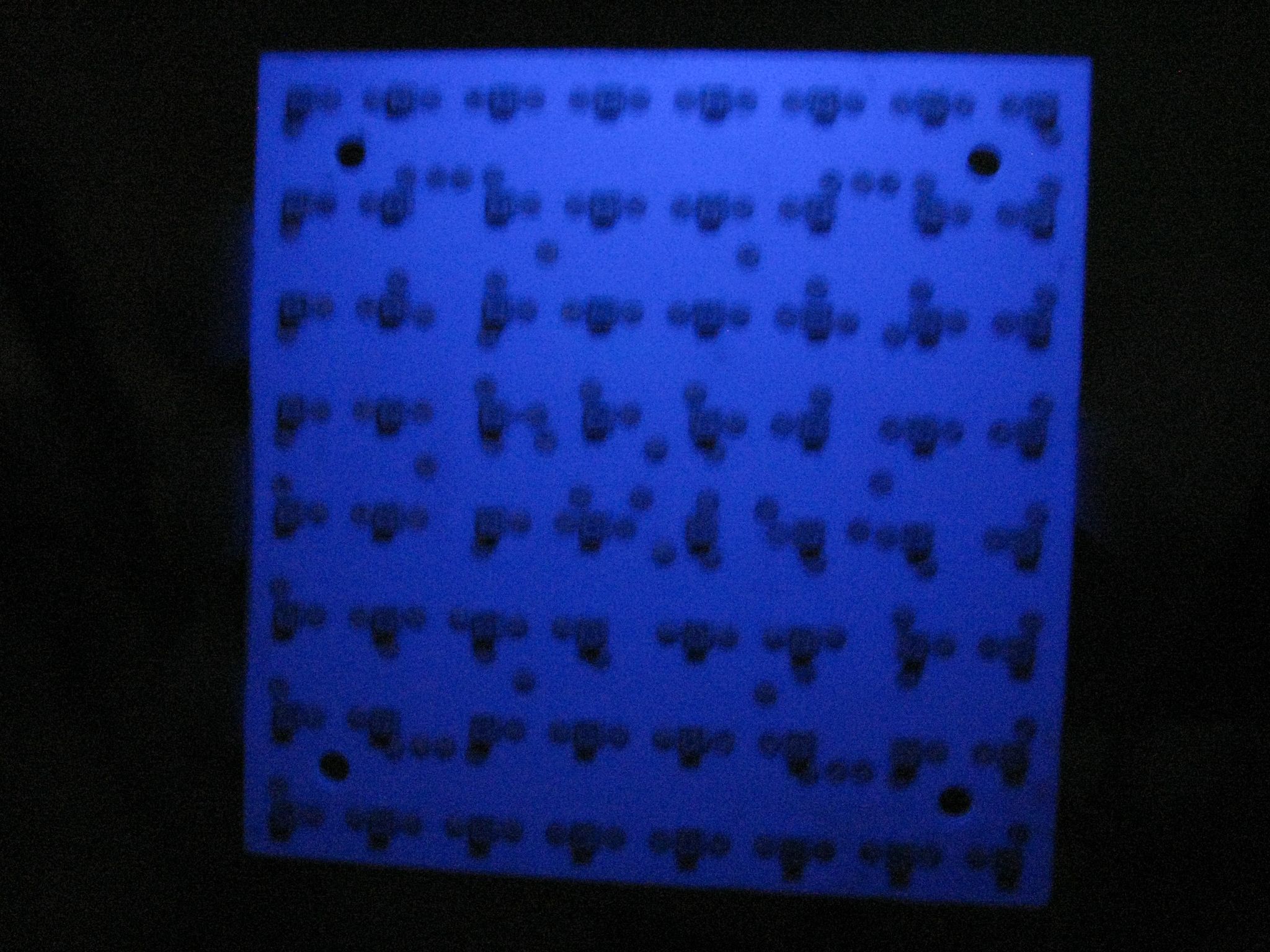}
  \end{center}
  \caption{Details of the \textit{CuFlon}$^{\circledR}$ boards and their mounting in NEXT-DEMO.}
  \label{fig:DICEs}
\end{figure}
The arguments discussed in section \ref{subsec:TOPO} led to the choice of a tracking plane 
for NEXT-DEMO instrumented with SiPMs at a sensor pitch of 1~cm. The plane consists of 256 Hamamatsu S10362-11-050P SiPMs distributed between 4 boards, each with 64 sensors (see figure~\ref{fig:DICEs}) and is positioned 2~mm behind the EL production region. The active area of a sensor is 1~mm$^2$ comprised of pixels of side 50~$\mu$m. Operating at a bias voltage of 73~V results in a gain of $\sim7.5x10^5$ at room temperature (precise calibration described in section~\ref{subsec:TrackCali}) and dark current at the level of 0.2-0.3 photoelectrons$~\mu$s$^{-1}$.

The SiPMs are mounted on multilayer \emph{CuFlon$^{\circledR}$} PCBs (PTFE substrate with copper layers, gold plated). Each board is supplied with one bias voltage via a FPC (Flat Printed Circuit) kapton cable which also extracts the sensor signals and the readings of a thermister positioned next to the board. In order to maintain the nominal supply voltage of the sensors and reduce detector dead time, 4 tantalum capacitors are connected to each board and supply the sensors with the required voltage after read-out.

\subsection{Front-End Electronics}
\label{subsec:FEE}
Signal processing, digitization and read-out is performed using an electronics chain comprised of a 16-channel front-end electronics board with 16 analogue paths, a digital processor and a front-end concentrator (FEC) which was developed by CERN and the NEXT Collaboration under the auspices of the RD51 read-out systems R\&D collaboration \cite{Toledo:2011zz}. Signals from the tracking plane first pass through a 3-stage analogue preprocessing before being digitized. In the first stage, the signals are subject to a preamplification and filtering in order to adapt them to the correct line impedance. At this stage, offset control can be used to optimize the dynamic range of the system. Signals are then passed through a gated integrator with a variable integration time. An integration interval of 1~$\mu$s was chosen for the data presented to limit the possibility of either saturating the ADC --- a saturation level of greater than 200~pe~$\mu\mbox{s}^{-1}$ is achieved --- or exceeding the maximum data-uplink rate. This interval results in a $z$ position resolution of approximately 1~mm --- considering a nominal drift velocity of 1~mm~$\mu\mbox{s}^{-1}$. The final analogue stage inverts the signal to match the ADC dynamic range.

Digitization of the signals is controlled via a small, low-power Xilinx Spartan-3 FPGA charged with the operation of the gated integrators, the building of a digitized frame via a 12-bit 1 MHz ADC and the transfer of these frames to the upstream read-out stage (the FEC). The SiPM front-end boards and the FEC are connected by means of a high-speed full duplex link consisting of two LVDS pairs in each direction over a standard CAT-6 network cable. The FEC sends a master clock to the front-end board FPGA over one of the LVDS pairs, thus providing system-level synchronization. A second downstream pair is used to send trigger and configuration commands to the front-end board. Data are sent to the DAQ PC via gigabit Ethernet link.

\subsection{Coating}
\label{subsec:coating}
Since the SiPMs are not directly sensitive at the emission wavelength of Xe (170~nm), it is necessary to use a wavelength shifting coating. TPB has been used since its emission properties match well with the sensitivity of the SiPMs and because it was used to coat the light tube. The SiPM PCBs (DICE boards) were coated to a TPB thickness of 0.1~mg~cm$^{-2}$ by evaporation under high vacuum using the techniques described in detail in~\cite{Alvarez:2012ub}. The peak of the emission spectrum of TPB (430~nm) matches the region of highest quantum efficiency of the SiPMs well and at the selected TPB thickness the transmittance of the shifted light is $> 96\%$ \cite{Alvarez:2012ub}. While higher transmittance is possible with a thinner coating, the variance of the effective conversion efficiency tends to increase. The thickness chosen is, as such, a good compromise between these two competing considerations.

\section{Calibration of the energy and tracking planes} 
\label{sec:Calibration}
In order to achieve a reliable, stable measurement of the energy
resolution of the detector, the photo-detectors must be calibrated so
that the number of photo-electrons detected by each sensor can be
accurately calculated from the digitized response. This calibration
must then be monitored during run time as the calibration constants
can be sensitive to external effects including the temperature and
pressure as well as to degradation of the sensors due to age or the
occasional sparks present in the TPC. A number of complementary
methods are used in NEXT-DEMO in order to achieve a reliable
calibration.

\subsection{Energy Plane Calibration} 
\label{subsec:Energy Plane Calibration}
\begin{figure}
  \begin{center}
    \includegraphics[width=0.65\textwidth]{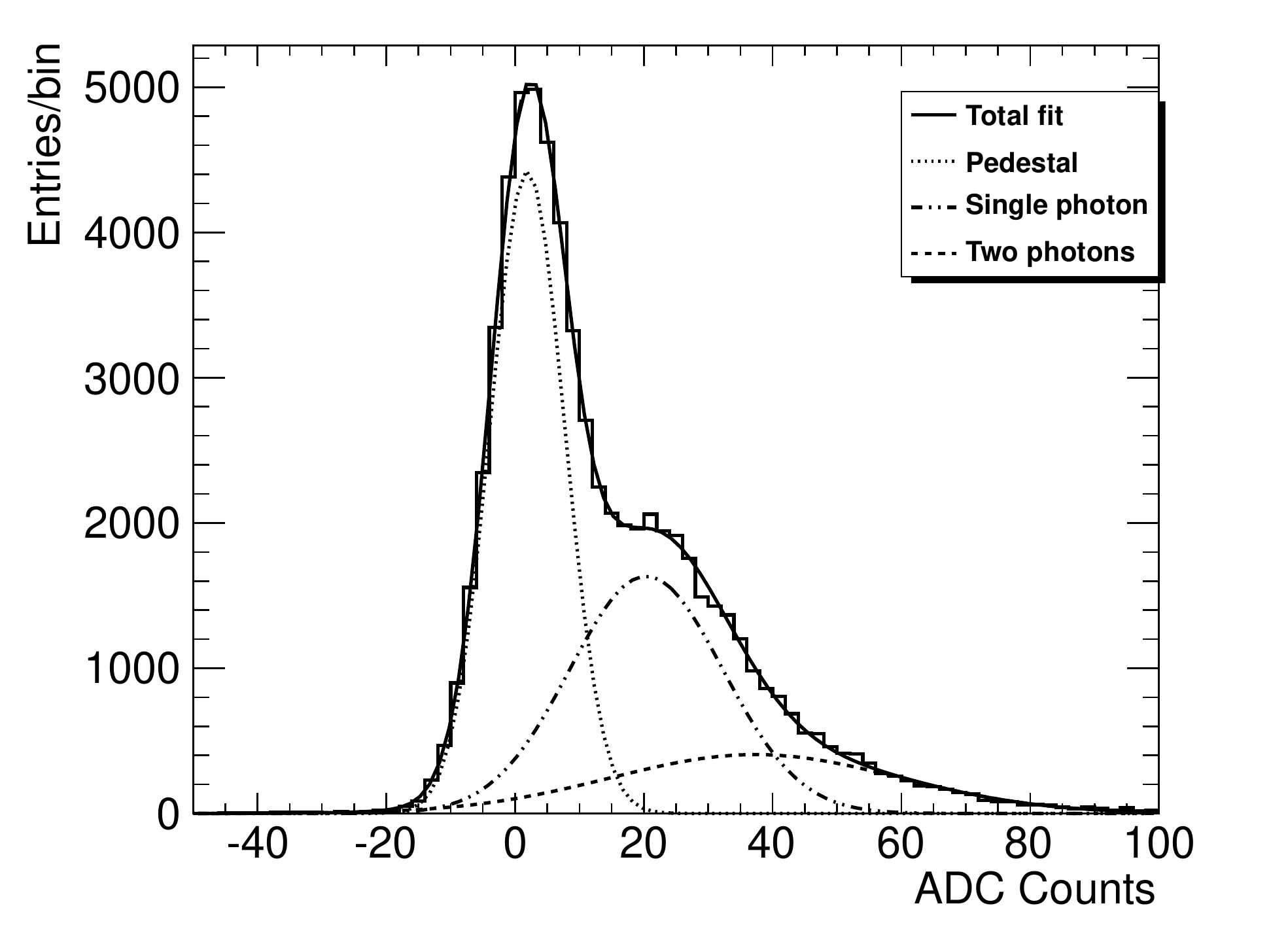} 
  \end{center}
  \caption{Typical single photo-electron spectrum of one of the
    energy plane PMTs. The single photon response is deconvoluted from
    pedestal and the two-photon response. }
  \label{fig:spe} 
\end{figure}

The PMTs of the energy plane are calibrated using their single photon
response. This is performed in situ using a 400~nm LED mounted in the
centre of the tracking plane. The PMTs are illuminated at a LED
intensity adjusted to limit the probability of the detection of more
than 1 photoelectron resulting in a spectrum of the form shown in
figure~\ref{fig:spe}. The distribution for each PMT
is then fitted using 3 Gaussian distributions representing the system
noise, single photon and two photon response of the PMT. The single
photon centroid in ADC counts represents the conversion factor between photoelectrons and
ADC counts with the two photon centroid giving an indication of the
robustness of the measurement by its relative position. The
calibration was repeated regularly during data taking both in
dedicated LED runs and taking advantage of the low rate of the
radioactive sources under study to take small groups of LED triggered
events every few hundred source triggers. In this way the stability of
the calibration could be monitored and any change in conditions
isolated to within a small set of events in the analyses.
%%%%%%%%%%%%%%%%%%%%%%%%%%%%
%%%%%%%%%%%%%%%%%%%%%%%%%%%%
The calibration of the PMTs has been found to be stable over the course
of data taking.

\subsection{Tracking Plane Calibration} 
\label{subsec:TrackCali}
 %%%%%%%%%%%%%%%%%%%%%%%%%%%%
\begin{figure}
\begin{center}
\includegraphics[width=0.65\textwidth]{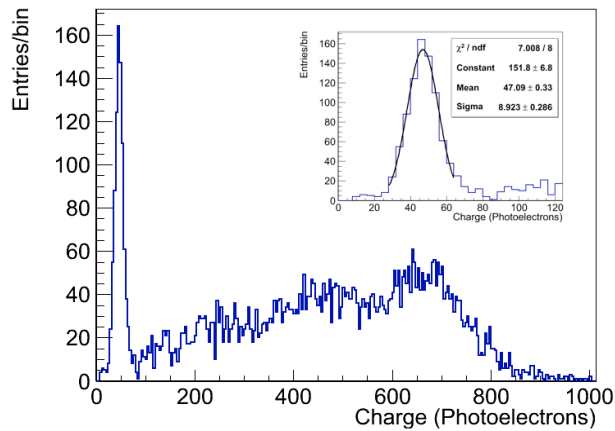}
\end{center}
\caption{``Low energy'' spectrum obtained with a \NA\ calibration
  source illuminating the detector. The x-ray peak and a gaussian fit
  for one of the SiPMs of the tracking plane are shown in the inset. The fit is performed for each SiPM in the plane.}
\label{fig:XR} 
\end{figure}

The sensitivity of SiPMs to atmospheric conditions is well documented
(see, for example, \cite{5873750}). For this reason and due to the method of bias of the
sensors (see section~\ref{sec:tracking}) it was necessary to perform a
pre-calibration of the sensors before mounting in the TPC. The absolute gain of
each sensor was measured using its single photon response to
illumination using a 400~nm LED as explained
in~\cite{Rodriguez:2012IEEE,Alvarez:2012xdb}. This allowed for the determination of
the appropriate distribution of voltage among the individual sensors
of a DICE and resulted in a gain spread of less than 4\% after these adjustments.

However, the read-out chain of the sensors once mounted in the
detector differs significantly from that of the external set-up. As
such, an independent calibration is required to give a true
representation of the relation between photoelectrons (pe) and ADC counts. A measurement of
the dark current of the SiPM channels can give a direct indication of
the conversion factor, however, the noise in the system means that the
peaks of the dark current distribution are not always obvious. For
this reason two independent techniques are used to perform an
equalisation of the channels. The response of the channels to the
X-ray peak present in source data and a photon transfer curve (PTC)
\cite{Janesick:2001} technique using a blue (400~nm) LED are presented here.

Gammas entering in the detector can interact with electrons in
the K, L shells of the xenon atoms. These interactions produce electrons
of $\sim$30~keV which, since they will travel only $\sim$0.6~mm in Xe
at 10~bar \cite{NISTESTAR}, can be considered point-like events in the
gas. Since the energy of these events is well defined they can be used
to equalize the SiPM response.

The equalization is performed by considering, for each source event,
only the SiPM with the largest observed charge. This results in
spectra of the form shown in figure~\ref{fig:XR} after pe
calculation using the externally determined constants. The x-ray peak is obvious at low energy and the contribution, in this
energy region, from non-x-ray events is limited since only the maximum
charge sensor is considered.

\begin{figure}
  \begin{center}
    $\begin{array}{cc}
      \includegraphics[width=0.495\textwidth]{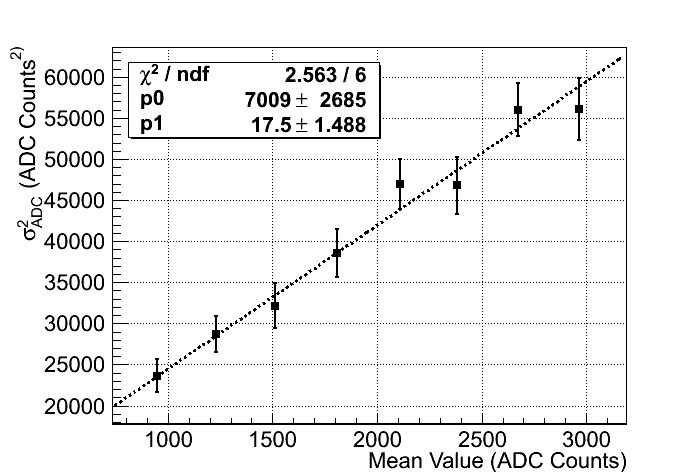} &
      \includegraphics[width=0.495\textwidth]{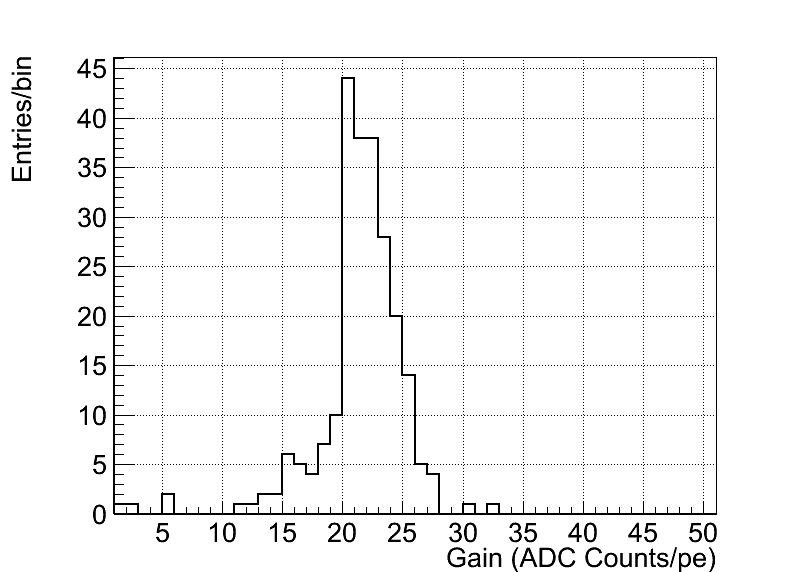}
    \end{array}$
    \caption{Extraction of conversion gain using the PTC
      method. Example linear fit (left) and the gains for all channels
    (right).}
    \label{fig:PTC}
  \end{center}
\end{figure}

While the x-ray method can be used over the course of a run with a
radioactive source to monitor the equalization of the gains, it is not
possible to measure the actual gain values using this method. As an
additional method of absolute determination of the gain in the absence
of the dark current method, the PTC~\cite{Janesick:2001} method using
a blue LED has been proposed. Recording multiple LED pulses at increasing
intensities, it can be demonstrated that there should exist a linear
region of a plot of variance versus mean intensity where the shot
noise of the arriving photons dominates over the other noise sources
and hence the gradient
is equivalent to the channel conversion gain:
%\begin{equation}
\begin{eqnarray}
  S_{ADC} = Gn_{pe}\, , \, \sigma_{ADC}^2 = (G\sigma_{pe})^2 \\
  \mbox{when photon stat. dominant }
  \sigma_{pe} = \sqrt{n_{pe}} \\
  \therefore \sigma_{ADC}^2 = GS_{ADC}
  \label{eq:PTC}
\end{eqnarray}
where $S_{ADC}$ is the digitized response, $n_{pe}$ is the number of
photoelectrons, $G$ is the conversion gain and, $\sigma_{ADC}$ and
$\sigma_{pe}$ are the standard deviations of the digitized signal and
photoelectrons respectively. Performing this measurement for each SiPM
channel individually, the conversion gains can be determined. As can
be seen in figure \ref{fig:PTC}-\emph{right panel} the gains, while
being dispersed to a greater extent than those measured outside the
chamber, are concentrated at values of $\sim$20. The few channels with
$G$ below 10 correspond to those which have been identified as being
faulty and are excluded from further analysis. For the channels
where dark current measurement is possible the gain measurements are found to be consistent.

\section{Event selection and position reconstruction}
\label{sec:select}
\subsection{Online trigger}
\label{subsec:ontrig}
\begin{figure}
  \centering
  \includegraphics[width=0.495\textwidth]{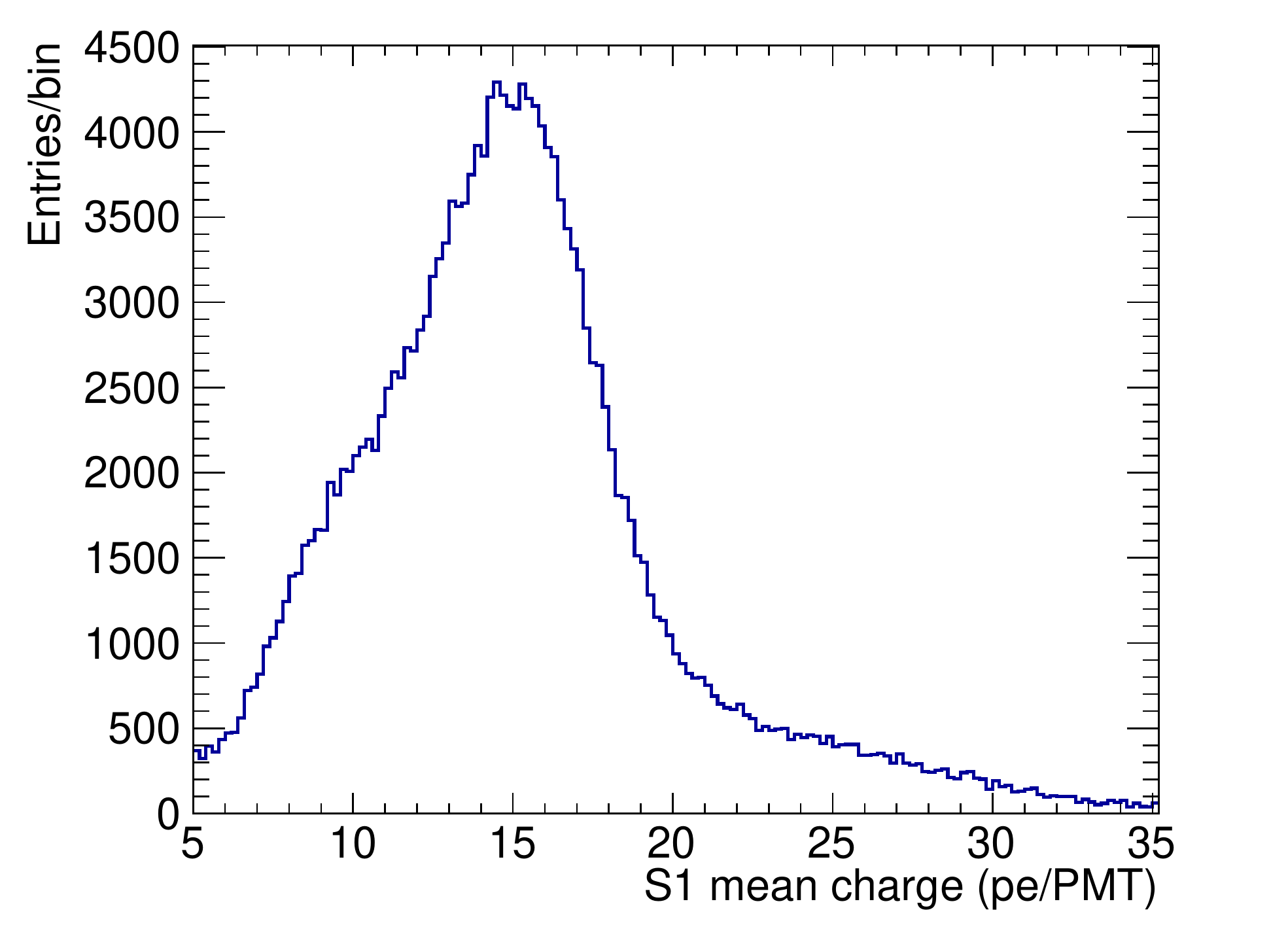}
  \includegraphics[width=0.495\textwidth]{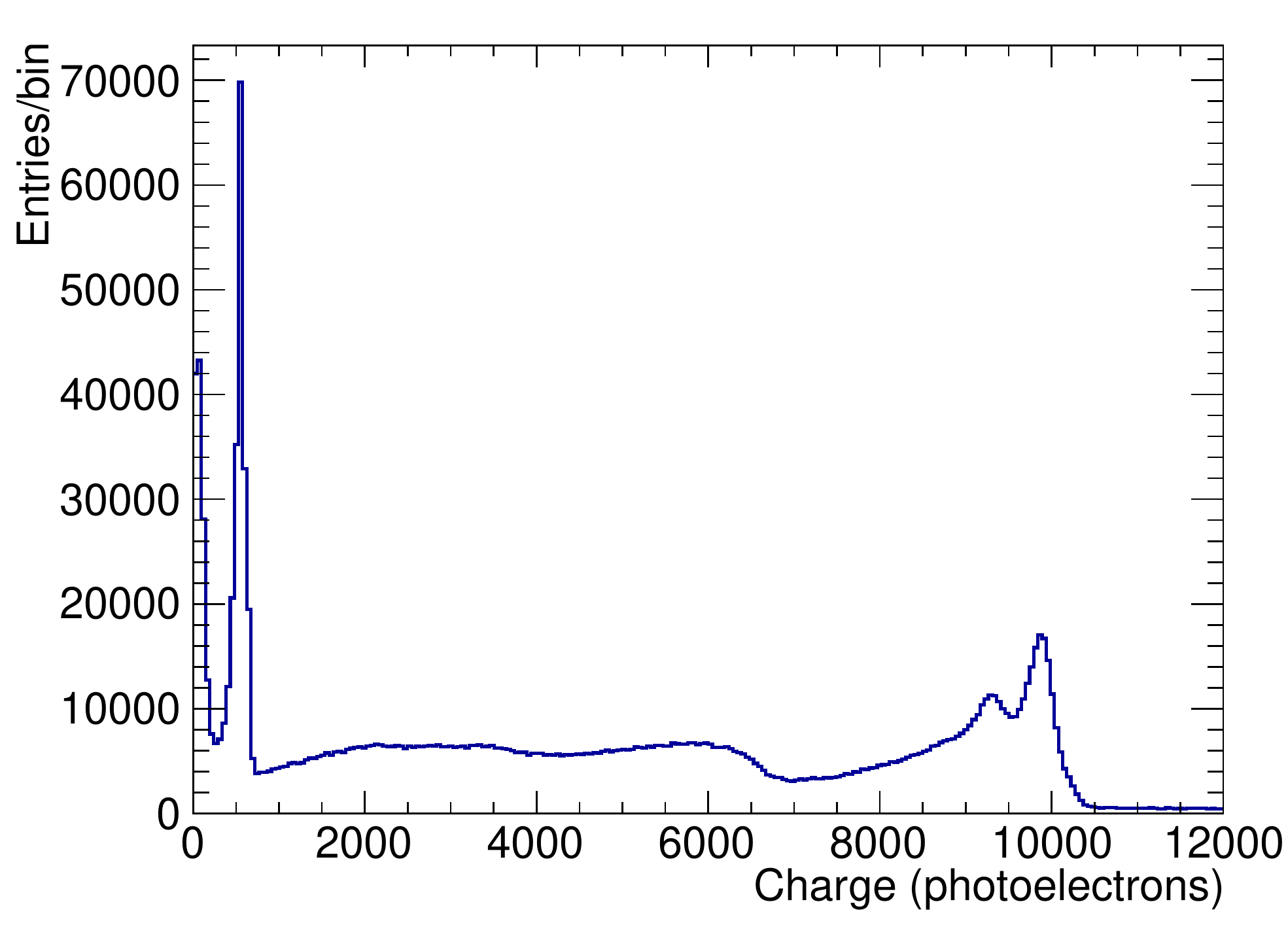}
  \caption{Primary scintillation spectrum (left) and the energy spectrum before any event selection (right).}
  \label{fig:RawE}
\end{figure}
Two distinct online triggers were used for the initial selection of the data presented in this paper. They used parameters based on the duration and amplitude of the scintillation (S1) signal and, for the first, coincidence between S1 and a signal in an NaI scintillator placed outside the source port. An S1 signal was defined as a pulse with between 1 and 20 photoelectrons per PMT with a duration shorter than 1~$\mu$s. Read-out was triggered in the first dataset when an S1 signal was detected within 25~ns of a pulse in the NaI scintillator. This trigger combination, possible with \NA\ due to the back-to-back photons produced in positron annihilation, ensures a pure source triggered dataset for analysis. In the second dataset, coincidence of S1-like signals was, instead, required between 3 of 4 central PMTs. The conditions for a positive trigger in each PMT were slightly more restrictive in order to ensure a trigger on S1 signal. This trigger configuration also has the possibility to be used under different data conditions where the external trigger would not be possible, for example with a $^{137}$Cs gamma source. The selection of cosmic muon events, however, requires both a larger integration window and greater maximum S1 charge.
Example spectra of S1 and S2 signals are shown in figure \ref{fig:RawE}.

\subsection{Pre-selection}
\label{subsec:presel}
An offline pre-selection of events is performed to further improve data quality prior to analysis. Events containing more than 1 S1-like pulse are rejected so that the event drift time and, hence, $z$ position in the detector, can be calculated unambiguously. While any number of S2-like signals are allowed per event, a minimum total S2 charge is required to eliminate low energy backgrounds and pulses generated in the cathode wires. The minimum charge is equivalent to 20 photoelectrons per PMT. Additionally, since light distribution in an event produced by a pulse in the cathode wires is non-uniform across the read-out plane, these events are rejected by demanding that the response of the individual PMTs not vary by more than 20\%.

\subsection{Position reconstruction and fiducial region}
\label{subsec:posRec}

\begin{figure}
  \begin{center}
    \includegraphics[width=0.65\textwidth]{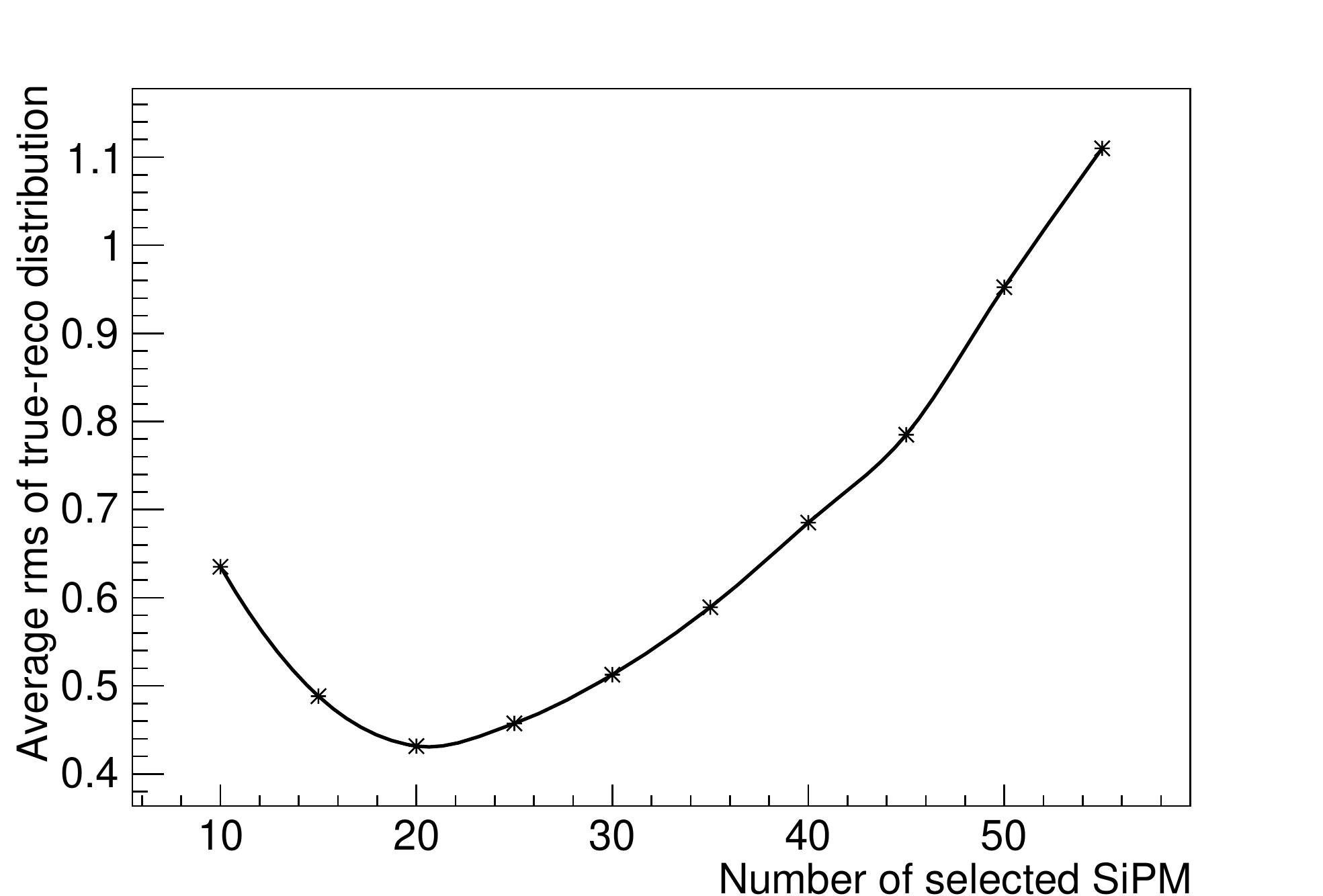}
  \end{center}
  \caption{Averaged difference in true and reconstructed $x$ position as a function of number of channels considered in the reconstruction for MC events.}
  \label{fig:BaryMC}
\end{figure}

While the $z$ position of the events in the chamber is unambiguously defined by the relative position in time of S1 and S2 signals and the drift velocity of the electrons in the gas, a number of possible methods can be used to reconstruct the $xy$ position of a charge deposit. For the data presented the barycentre method is used. Monte Carlo studies of the detector geometry indicate that this method is optimal using the signals from only the $\sim$20 SiPMs with greatest charge (figure~\ref{fig:BaryMC}). The analyses presented here make use of this observation.

In order to eliminate events that are not fully contained in the field cage and those expected to be reconstructed with greater error, a fiducial volume based on the reconstructed $xy$ position must be defined. \NA\ gamma induced electrons can travel between a few hundred microns and a few cm, however, due to multiple scattering the most probable extent of the higher energy events is a few mm. Thus, events reconstructed within a few mm of the outer edge of the tracking plane can be expected to have high probability to have lost some of their charge to the outer walls of the detector. These observations inspired the definition of a fiducial cut which excludes those events with a reconstructed mean $|x|$ ($|y|$) position greater than that of the sensors in the penultimate column (row) of the plane. This cut defines a fiducial region that covers 77\% of the total tracking plane active area. Moreover, to avoid events very close to the cathode which could lose charge and to exclude the region with low statistics due to the position of the source port, an additional requirement that events be reconstructed with $100 \leq z \leq 240~\mbox{mm}$ was implemented. As a result the fiducial volume used in the analyses presented is a cuboid of volume $120\times 120\times 140~\mbox{mm}^3$ which corresponds to 36\% of the active region of the TPC, 3.7 times larger than that used in the previous analysis~\cite{Alvarez:2012xda}. A similar definition would result in a fiducial volume corresponding to 88\% of the active region of a TPC of the size of NEXT-100.

\subsection{Event topology}
\label{subsec:evtTop}
\begin{figure}
  \begin{center}
    \includegraphics[width=0.495\textwidth]{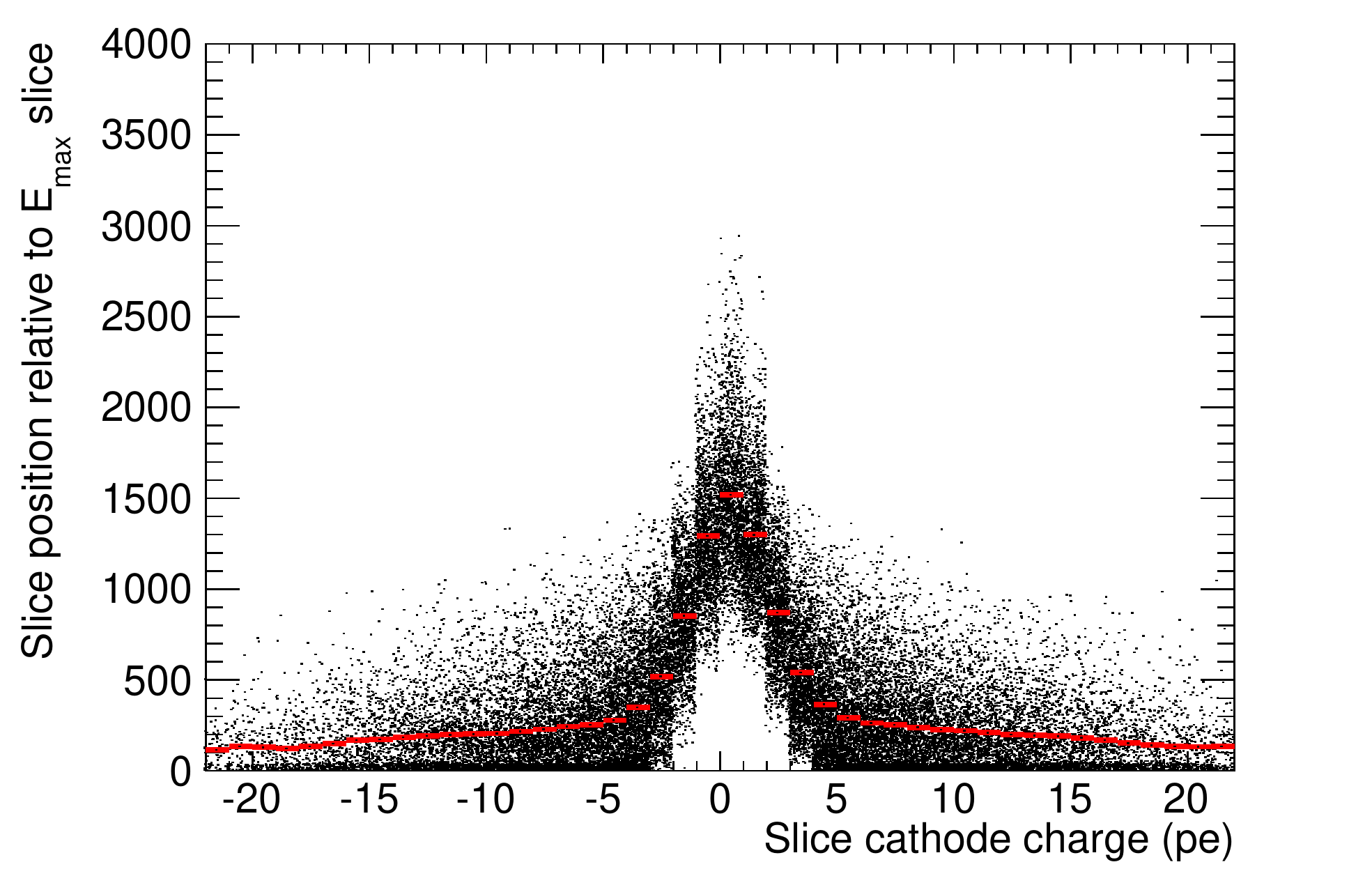}
    \includegraphics[width=0.495\textwidth]{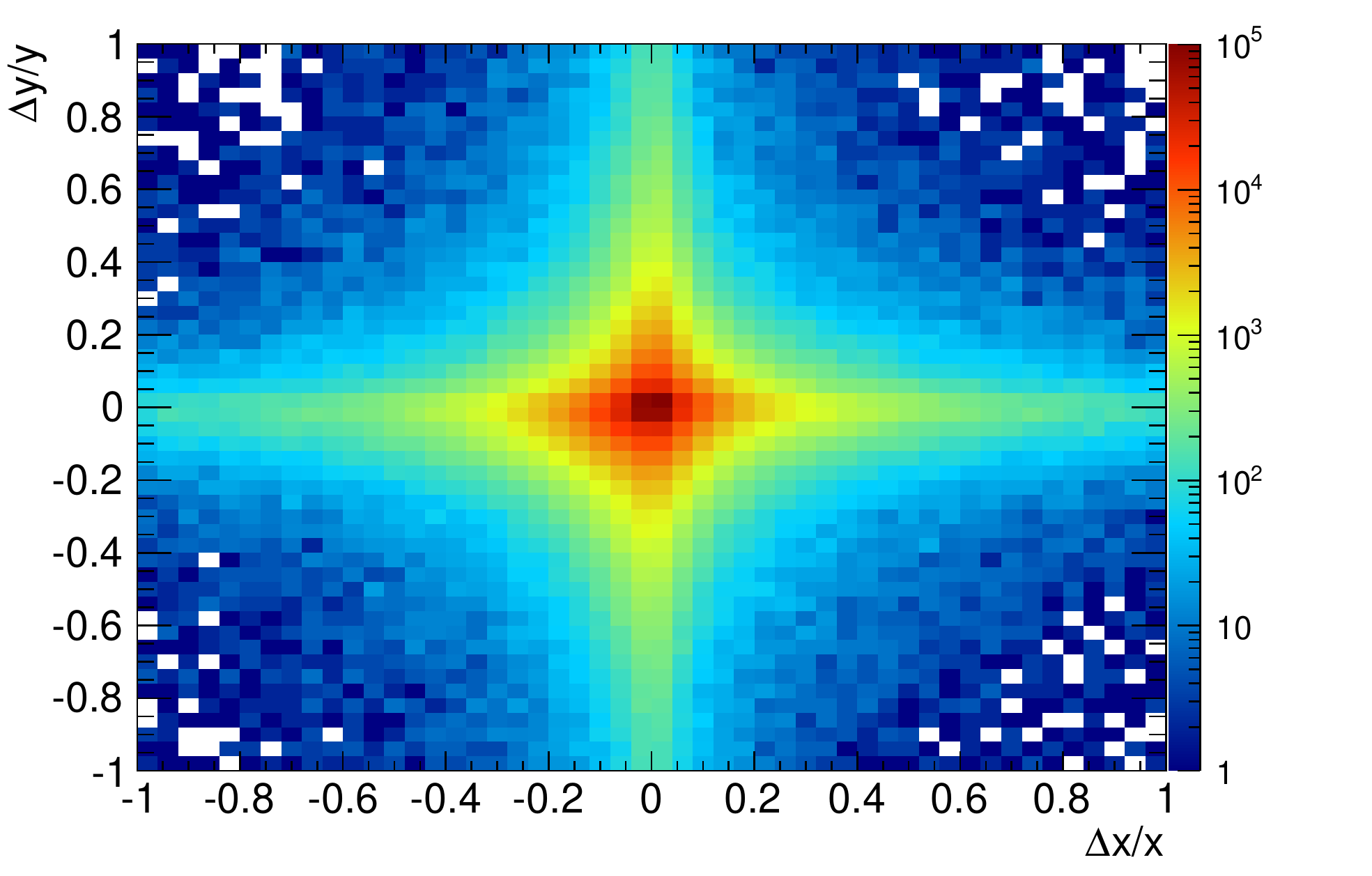}
    \caption{Slice position relative to the maximum (left) and the pull on the ``blob'' position compared to the barycentre of the event as a whole ($\Delta  y/y~(\Delta x/x)$ being the difference normalised to the total charge position) (right).}
    \label{fig:blob}
  \end{center}
\end{figure}

Using the event timing and the barycentre calculated using the total integrated signal in each SiPM, it is possible to obtain an estimation of the position of the energy deposit associated to the interaction of a particle. This definition is a good approximation when the particle behaviour is point-like, as is the case for x-rays. However, the electrons produced by the interactions of the 511~keV photons emanating from the \NA\ calibration source are not necessarily point-like. They have finite probability to travel several centimetres in the gas. As can be seen in figure \ref{fig:blob}-\emph{left panel}, the majority of the charge in \NA\ events tends to be concentrated within 4~$\mu$s either side of the time sample ($\mu$s width slice) with maximum charge. Defining this region as the ``blob'' and calculating the position of this deposit compared to that calculated from the full integrated charge of the event, it can be seen (figure \ref{fig:blob}-\emph{right panel}) that the global position is still a good estimator of the position of the energy deposit. 

A more involved reconstruction is required to understand the actual topology of individual events. A first approximation of the event topology can be made by subdividing, to a minimum width of 1 time sample, the charge in time. This allows for a deeper understanding of an event's $z$ structure and is the starting point for track reconstruction.
\begin{figure}
  \begin{center}
  \includegraphics[width=0.495\textwidth]{img/OnlySiPM.pdf}
  \includegraphics[width=0.495\textwidth]{img/Points.pdf}
 \includegraphics[width=0.495\textwidth]{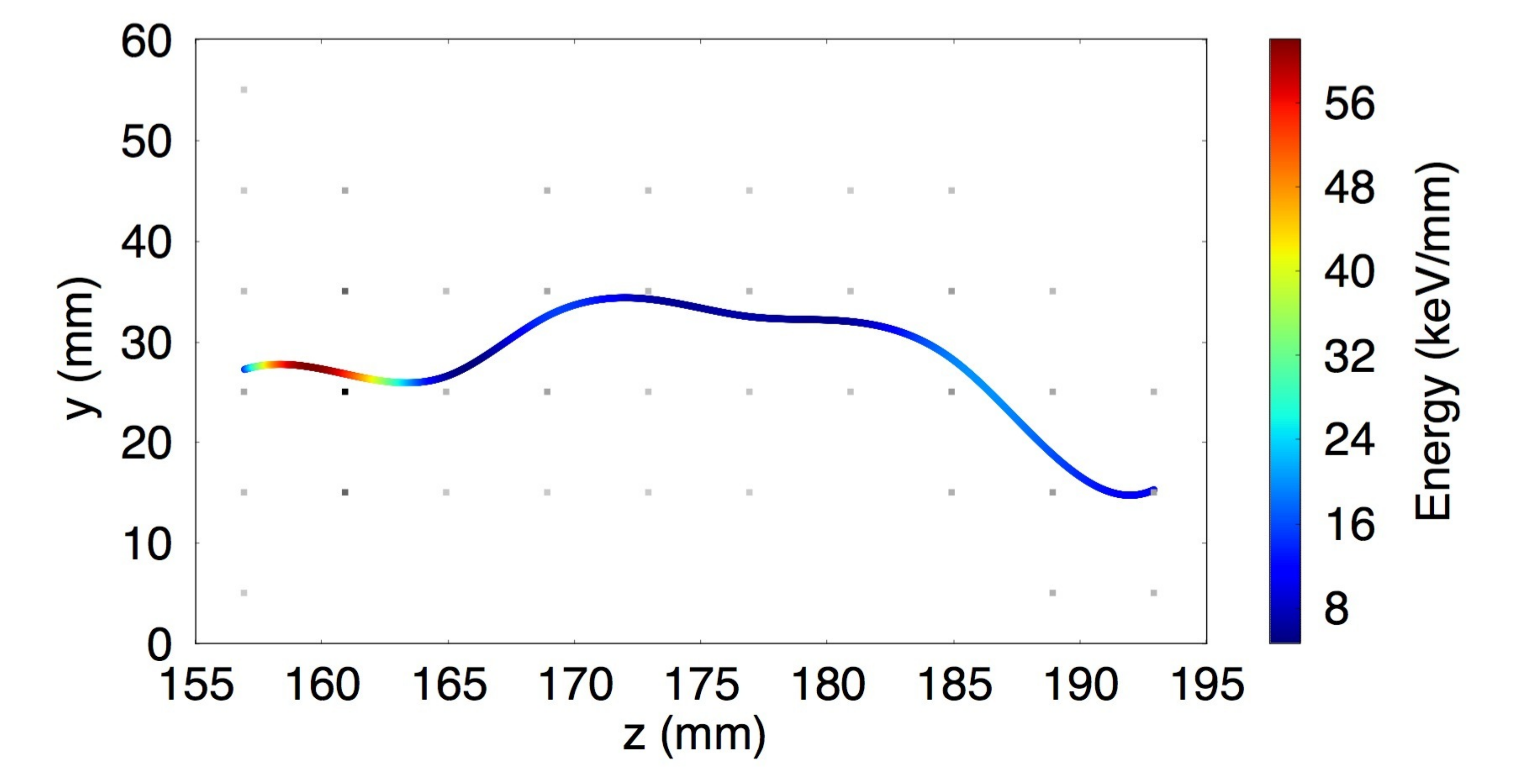}
  \includegraphics[width=0.495\textwidth]{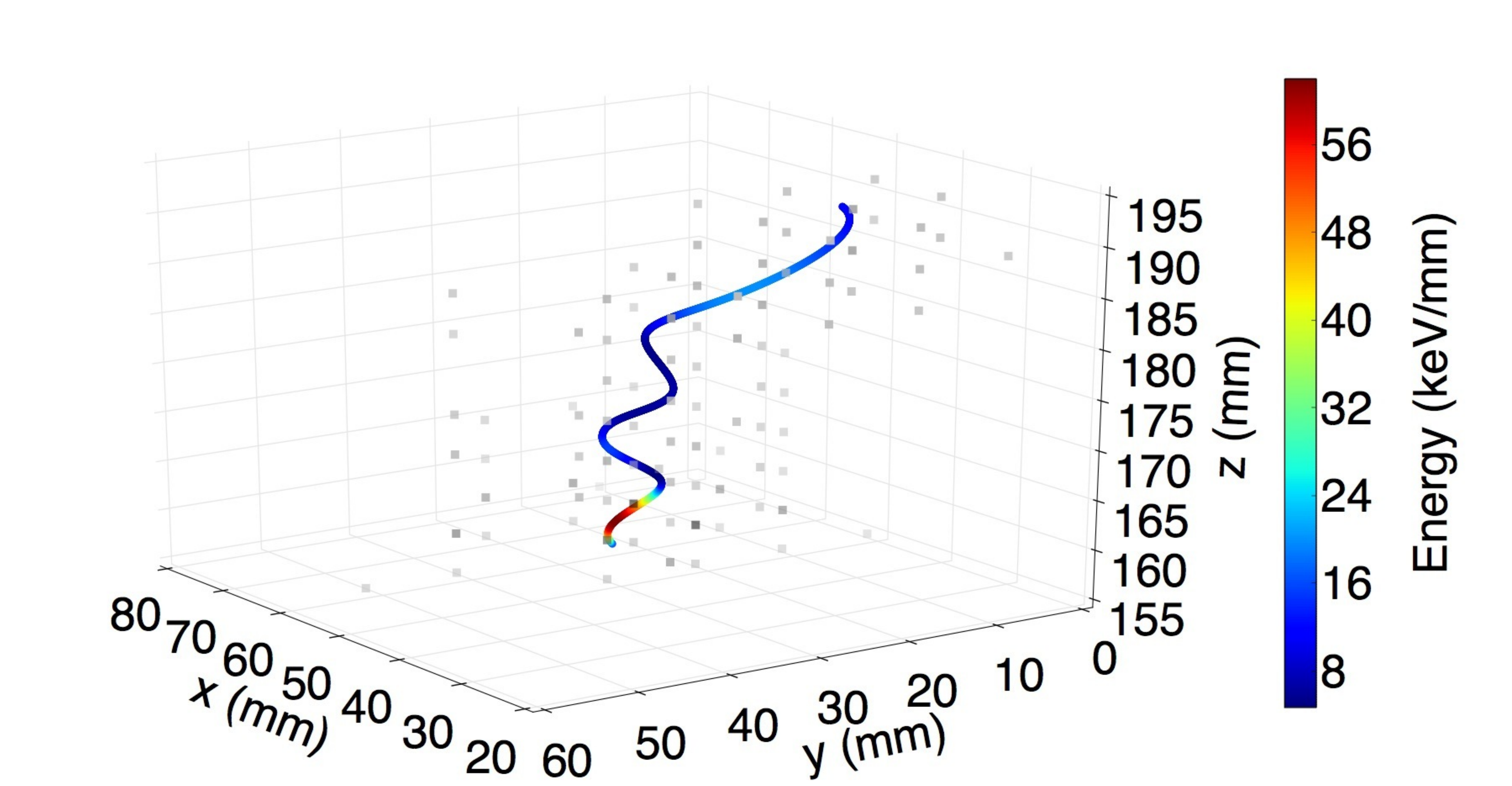}
\end{center}
  \caption{Example of \CS\ track reconstruction: The charge of the different SiPMs is split into slices of 4~mm width in $z$ (top left). One point is calculated for each slice using the barycentre method and the energy of the points is then associated with the measurement made in the cathode (top right). A cubic spline is used to interconect the different points; YZ projection (bottom left) 3D (bottom right). }
  \label{fig:Track}
\end{figure}
For the analysis presented here, slices of 4~$\mu$s width were used, on the grounds that this width is comparable to the width of the blob, as defined above, and the time an electron needs to cross the EL region ($\sim$3~$\mu$s). In addition, the charge collected in a 4~$\mu$s width window is sufficient to achieve a reliable \textit{xy} reconstruction.  Each time slice, while having a single $z$ coordinate, can be comprised of multiple $x,y$ points. In this preliminary analysis a single $xy$ point is reconstruced per slice and those events exhibiting slices with more than one isolated charge deposit in the tracking plane are excluded from analysis until a more sophisticated clustering algorithm is developed.

The $xy$ position of a slice is reconstructed using its barycentre. This position then has the energy recorded in the cathode for the same time interval associated with it so that the event $dE/dz$ can be studied. The energy and position information are then used to calculate a cubic spline between the individual points in order to obtain a finer description of the path. 

\begin{figure}
  \begin{center}
    \includegraphics[width=0.495\textwidth]{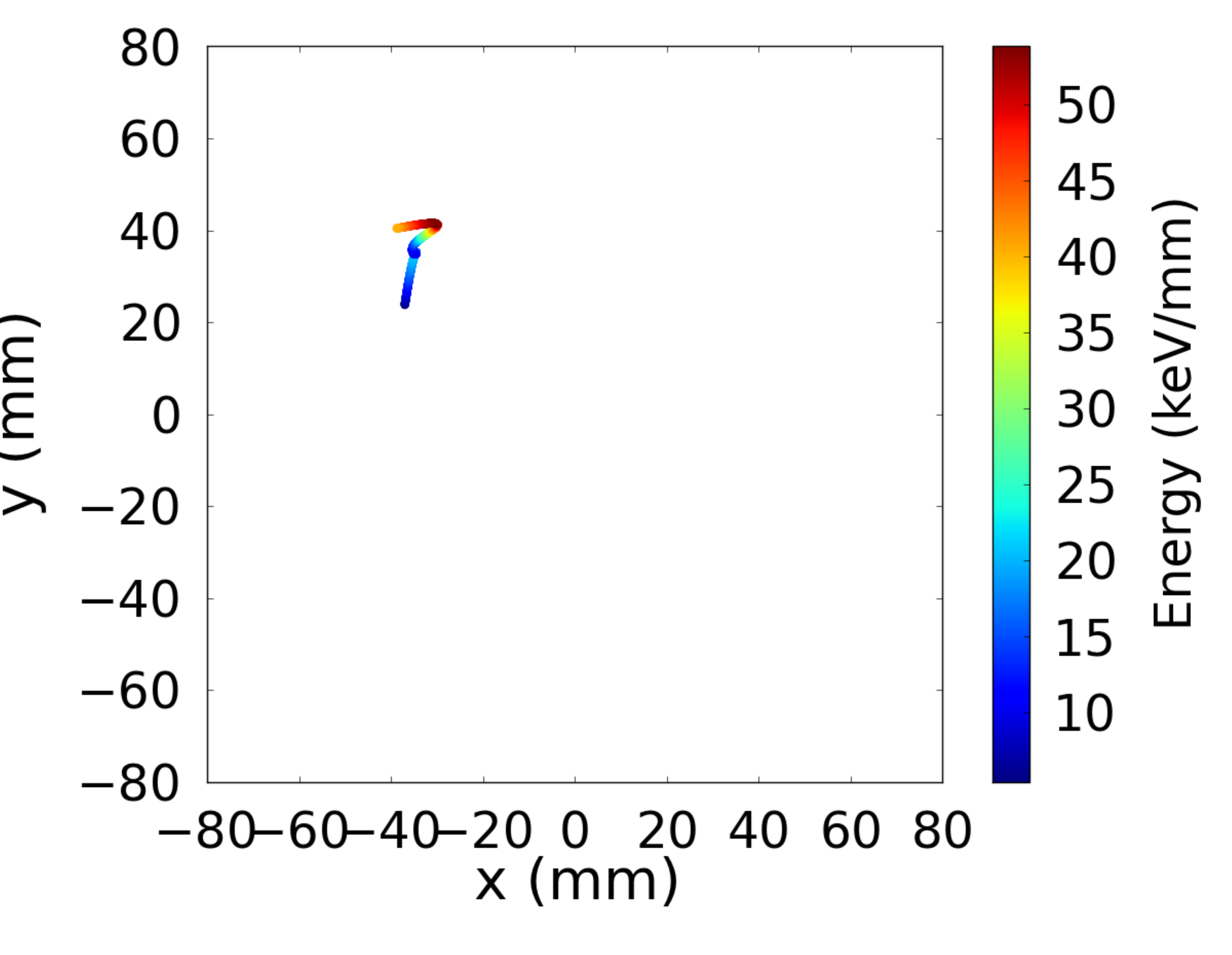}
    \includegraphics[width=0.495\textwidth]{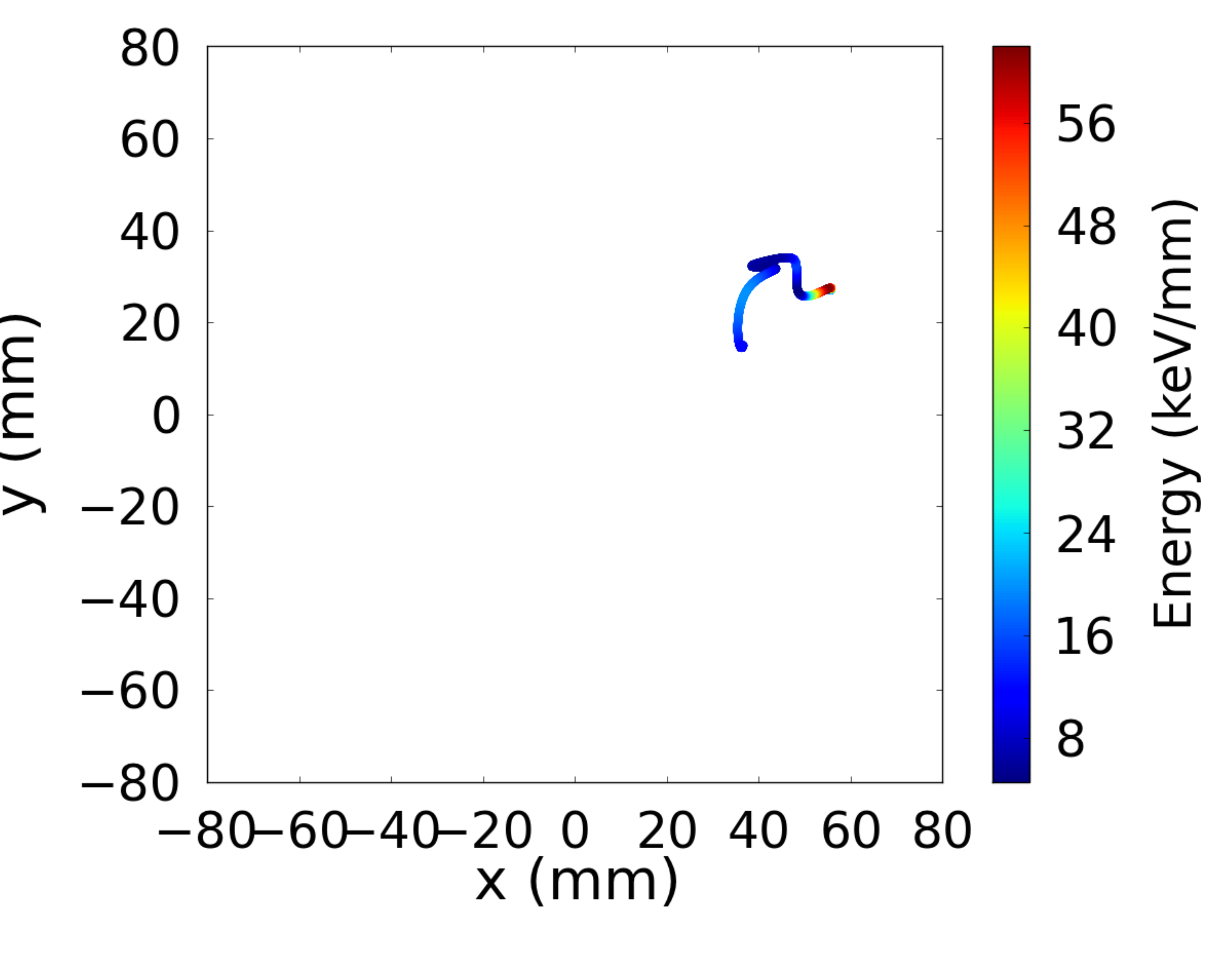}
    \includegraphics[width=0.495\textwidth]{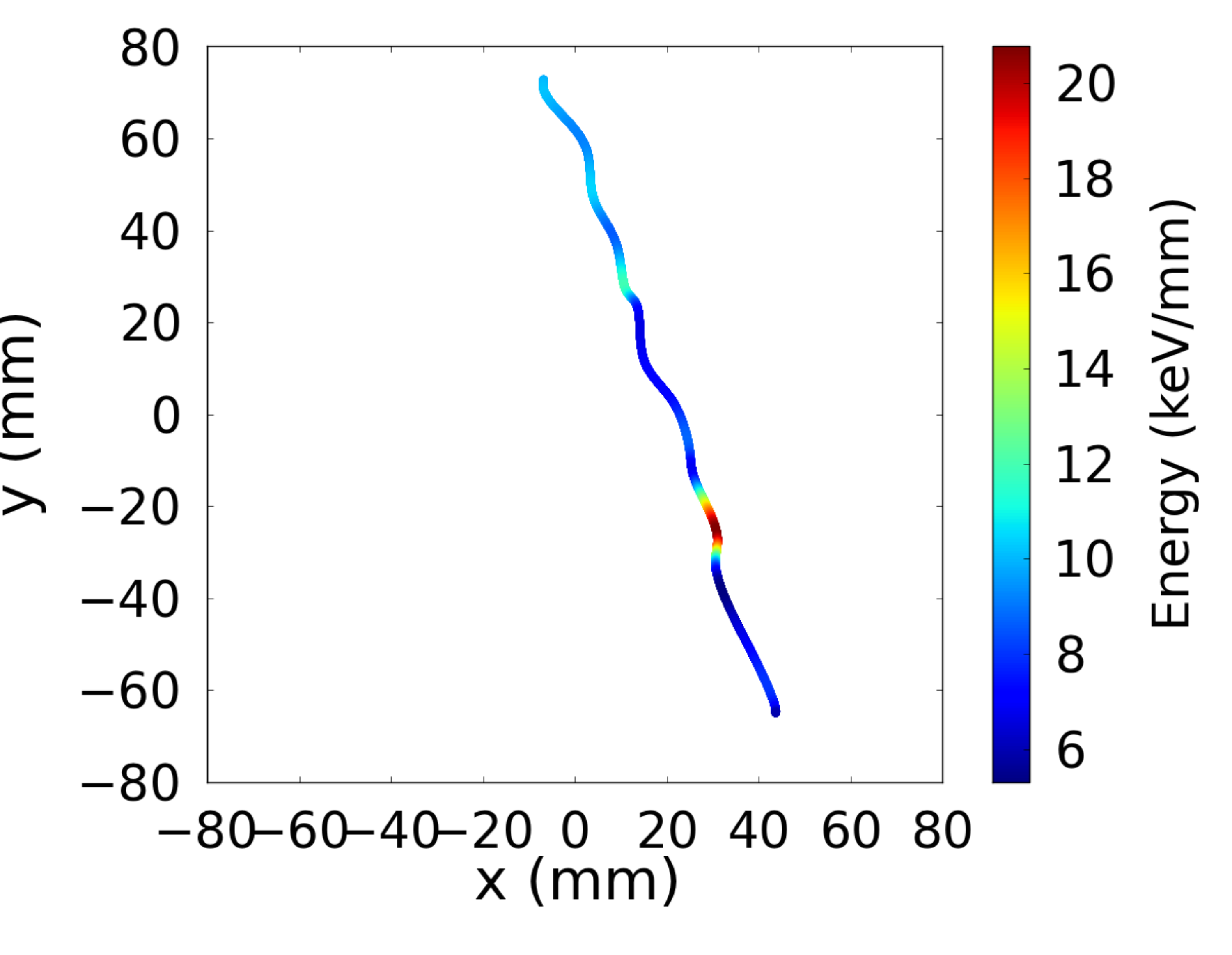}
    \caption{Examples of \NA\ (top left), \CS\ (top right) and muon (bottom) track $xy$ plane projections. Tracks reconstructed from NEXT-DEMO data.}
    \label{fig:TrackExs}
  \end{center}
\end{figure}

Figure~\ref{fig:Track} illustrates the method applied to the reconstruction of a \CS\ photoelectric event. The reconstructed electron trajectory presents all the features found in the Monte Carlo: a tortuous path due to multiple scattering, a ``wire'' region of MIP deposition and a  blob of high energy deposit towards its end. In figure~\ref{fig:TrackExs} examples of three different tracks are shown with the same scale. As expected the \NA\ track is smaller than the track from \CS. It is also notable that in the electron tracks the end-point is clearly visible but this is not the case for the muon.

\section{Energy reconstruction}
\label{sec:engRes}
NEXT-DEMO must prove that energy reconstruction can be performed with a resolution
compatible with extrapolation to 1\% FWHM at \Qbb\ (the target value in
the NEXT-100 TDR) or better over a large
fiducial volume. The most important
factors affecting the energy resolution of the NEXT-DEMO detector are
geometrical inhomogeneities in response and any time variation of the
detector gain, both in terms of EL yield and sensor
calibration~\cite{Alvarez:2012xda}. 

\subsection{Charge time dependence}
\label{subsec:Qtime}
As described in section~\ref{sec:Calibration} the calibration constants
of the photodetectors were constantly monitored over the course of
data taking so that any variation could be taken into
account. Additionally, the temperature and pressure in the TPC have
been monitored allowing for the study of the correlation between these physical
factors and the detector response. 
\begin{figure}
  \begin{center}
      \includegraphics[width=0.495\textwidth]{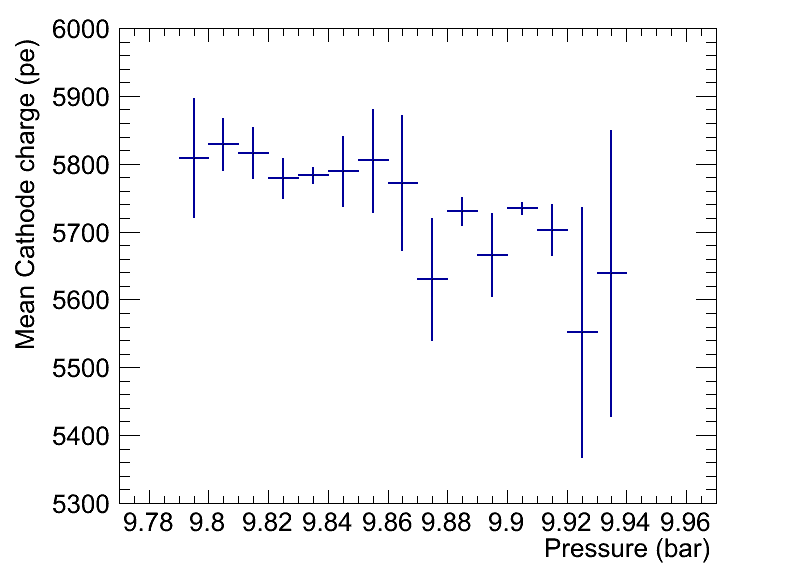}
      \includegraphics[width=0.495\textwidth]{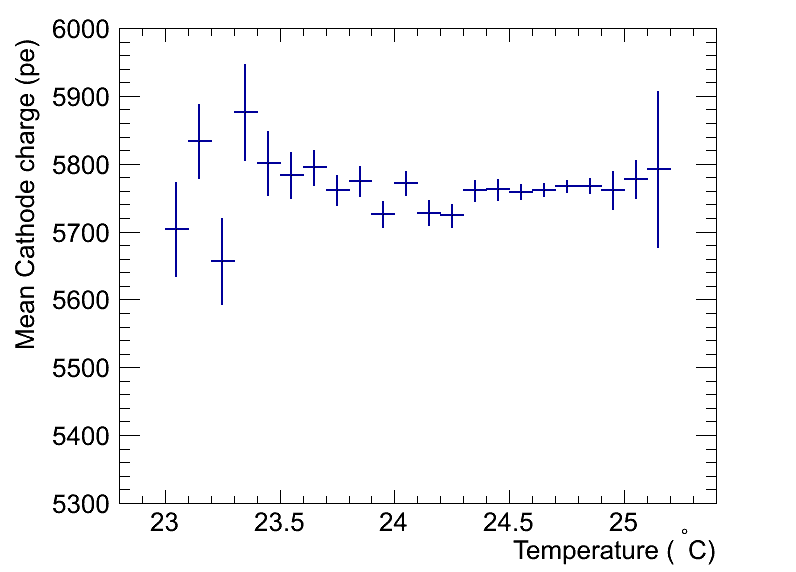}
    \caption{Relation of observed cathode charge with ambient
      conditions: Vessel pressure (left) and vessel temperature
      (right).}
    \label{fig:ambient}
  \end{center}
\end{figure}
A correlation between the measured pressure and charge is observed in
the data as can be seen in figure~\ref{fig:ambient}-\emph{left
  panel}. The temperature of the vessel has a significantly
smaller effect on the PMT charge. Both temperature and pressure oscillate at the level of
$\sim$0.2\% every 10 minutes due to the cycle of the hot getter which
purifies the gas and by up to 3\% due to air conditioning and activity
in the laboratory over a timescale of 24~hours. Other possible
sources of time dependent variation are the occurrence of sparks in the
TPC and variation of the gas purity. The increased light yield due to the TPB-coated tracking plane (see
section~\ref{sec:tracking}) enabled the use of lower EL fields
compared to previous analyses, and
improved gas isolation and recirculation reduced the
probability of sparking and of electron attachment due to
impurities. By monitoring these variables, the data can be gain
corrected to improve consistency and, in the case of any major changes
in conditions, affected data could be removed from analysis.

\subsection{Geometrical correction}
\label{subsec:geocor}
\begin{figure}
\centering
\includegraphics[width=0.495\textwidth]{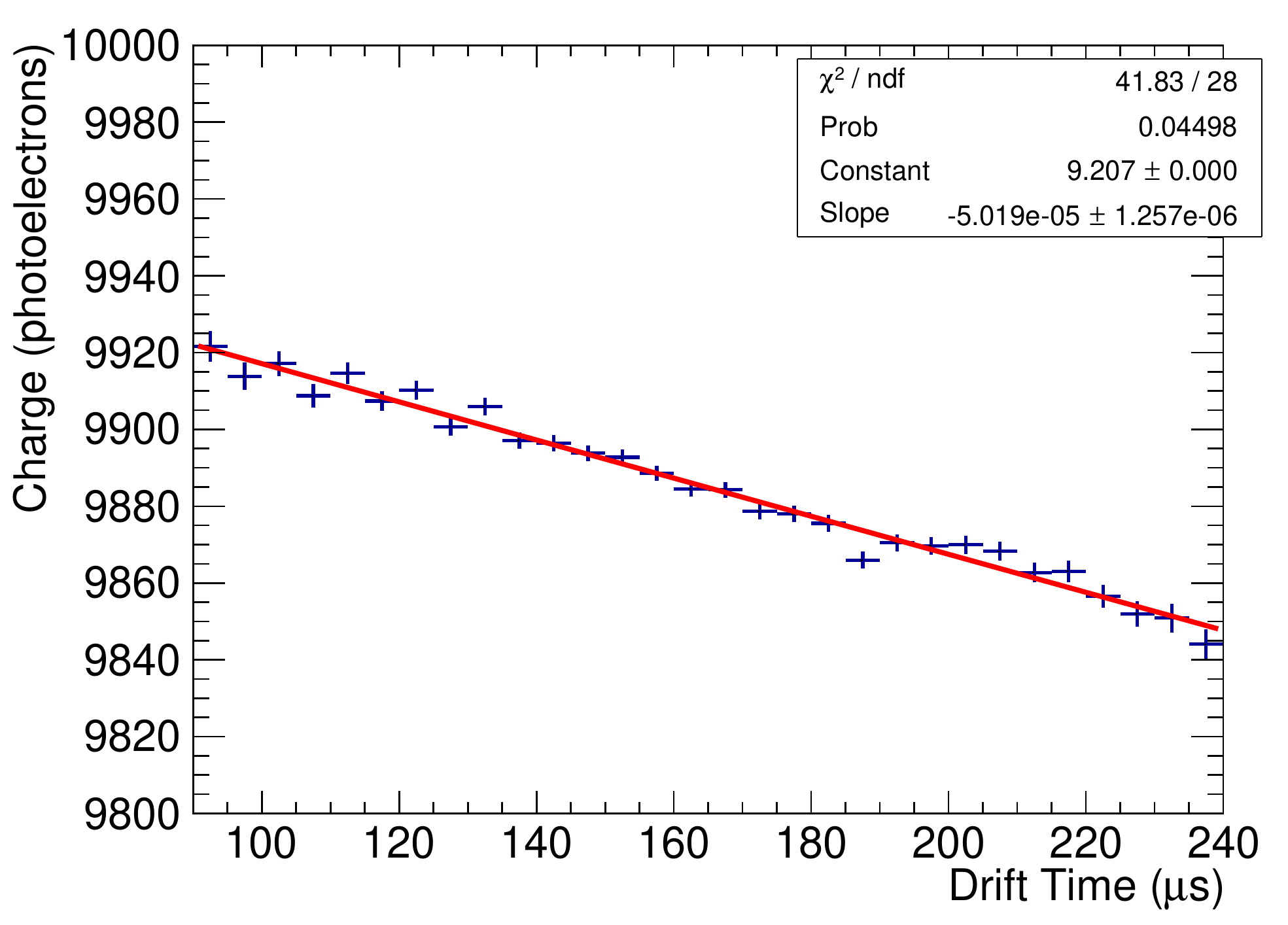}
\includegraphics[height=5.cm]{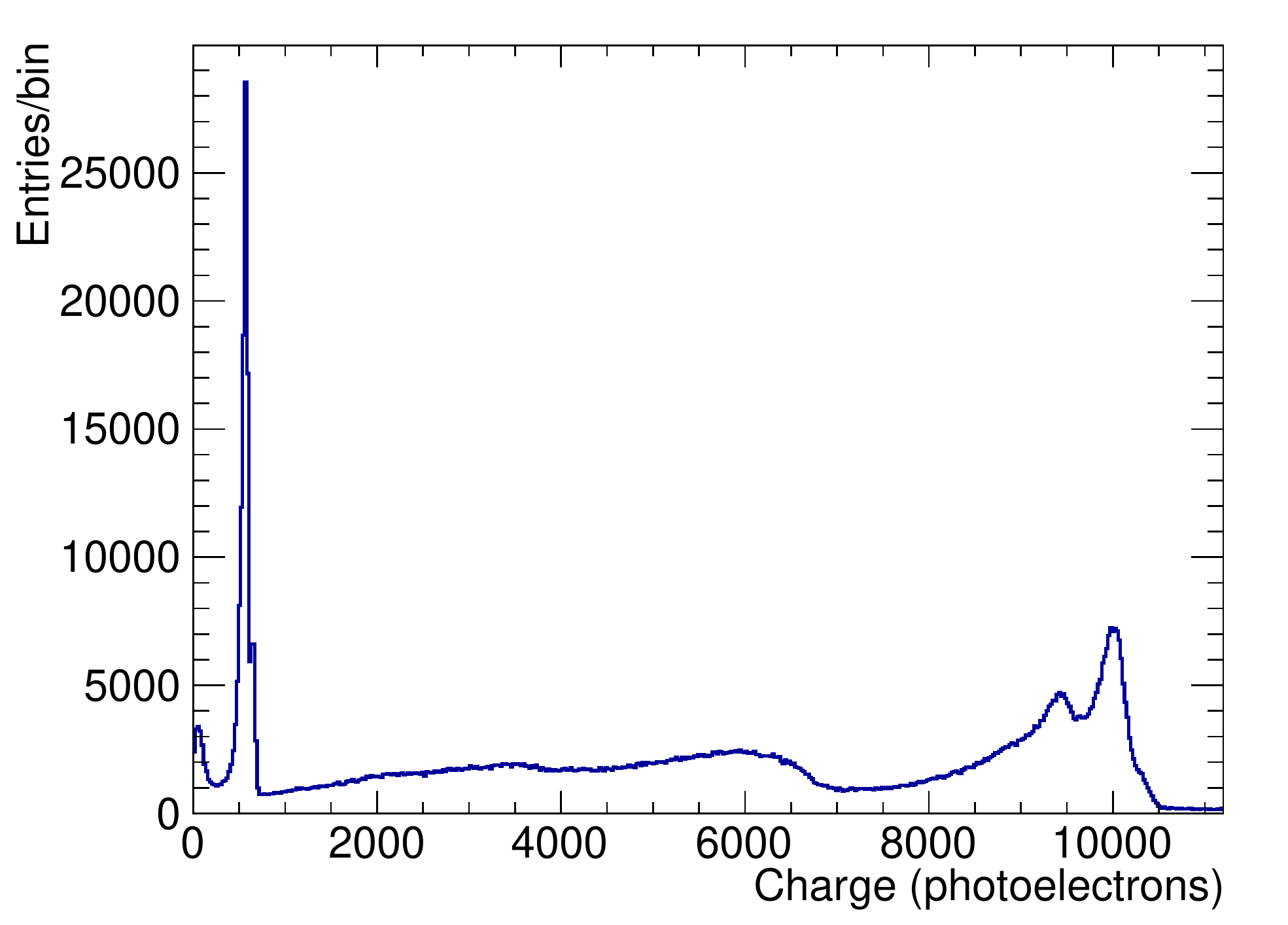}
\caption{Energy dependence with drift time due to electron attachment
  in the gas (left) and the energy spectrum after correction (right).}
\label{fig:zcorr}
\end{figure}

\begin{figure}
  \begin{center}
    \includegraphics[width=0.495\textwidth]{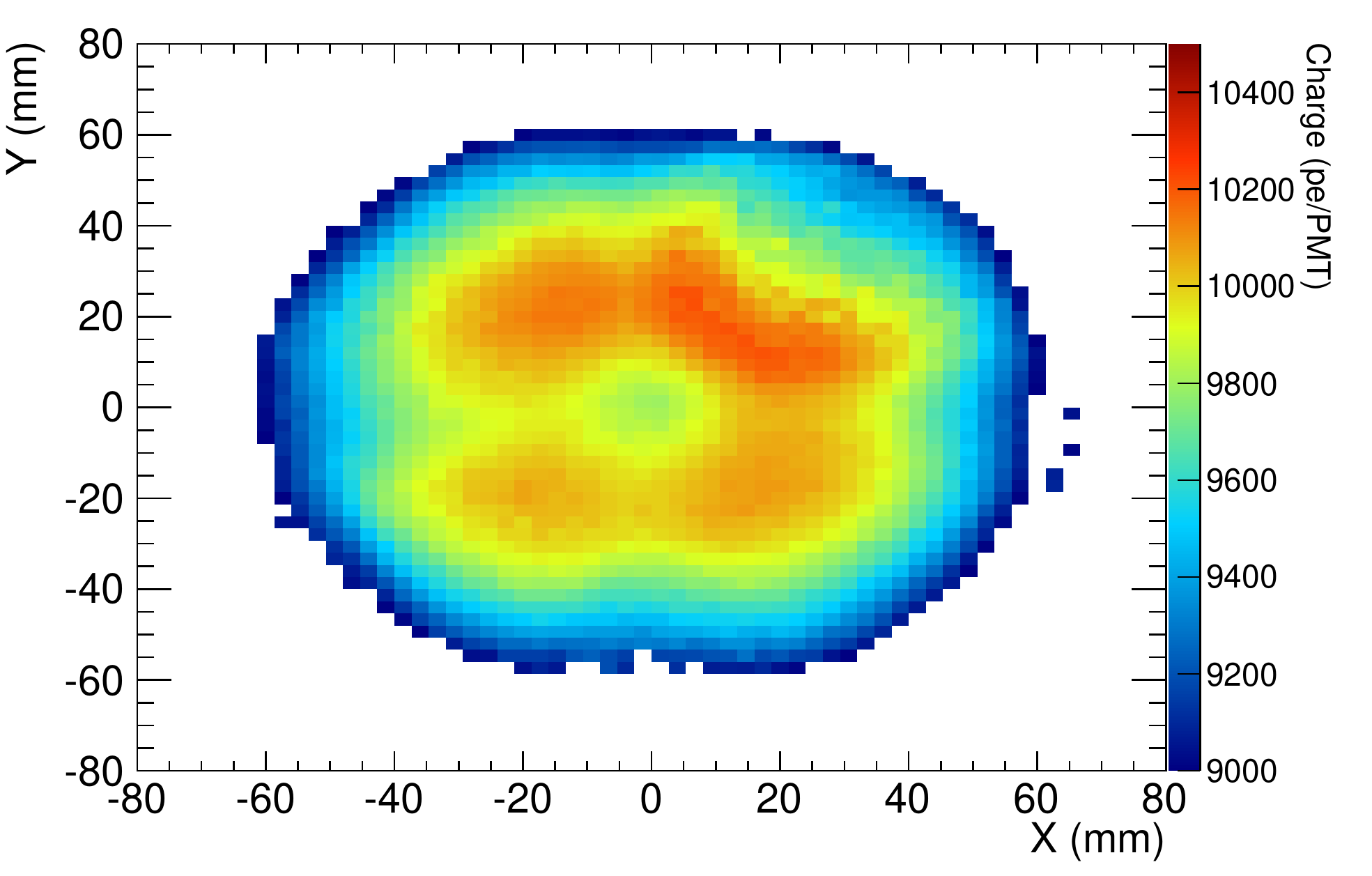}
    \includegraphics[width=0.495\textwidth]{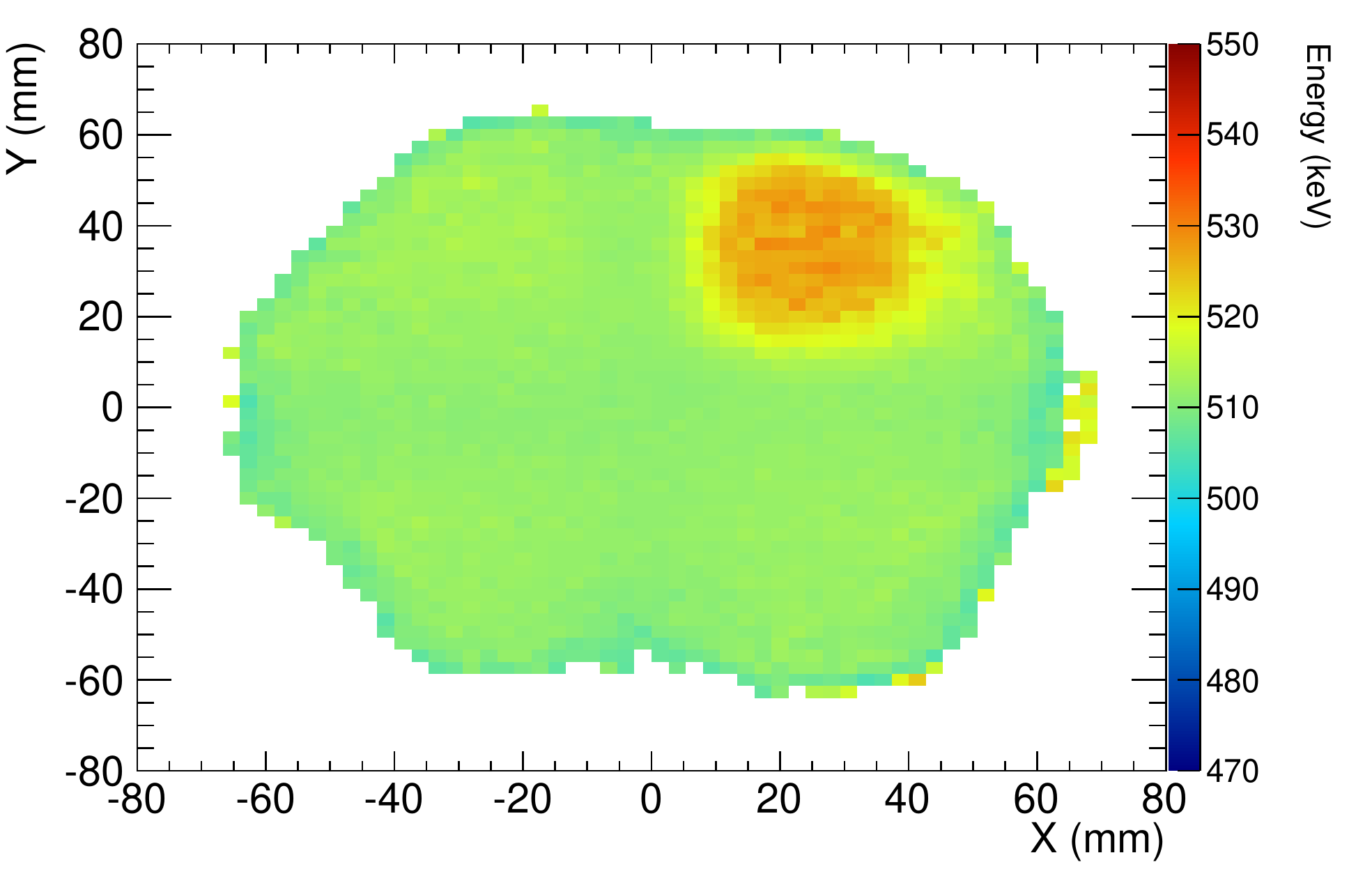}
    \includegraphics[width=0.495\textwidth]{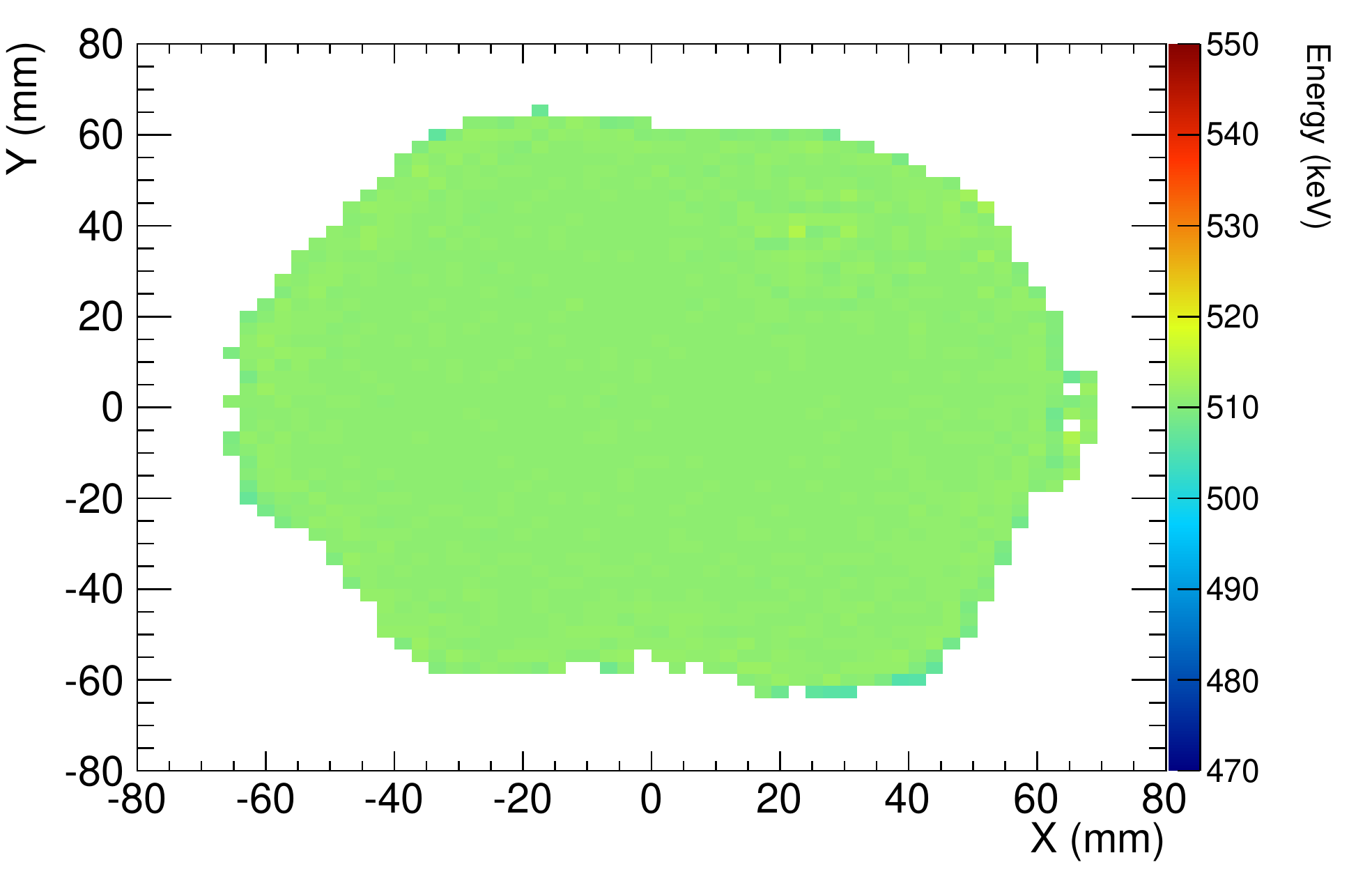}
  \end{center}
  \caption{Top Left: Profile before $xy$ corrections. Top Right: Profile after first
    correction. Bottom: Profile once convergence is achieved.}
  \label{fig:xyprofiles}
\end{figure}

\begin{figure}
\centering
\includegraphics[width=0.495\textwidth]{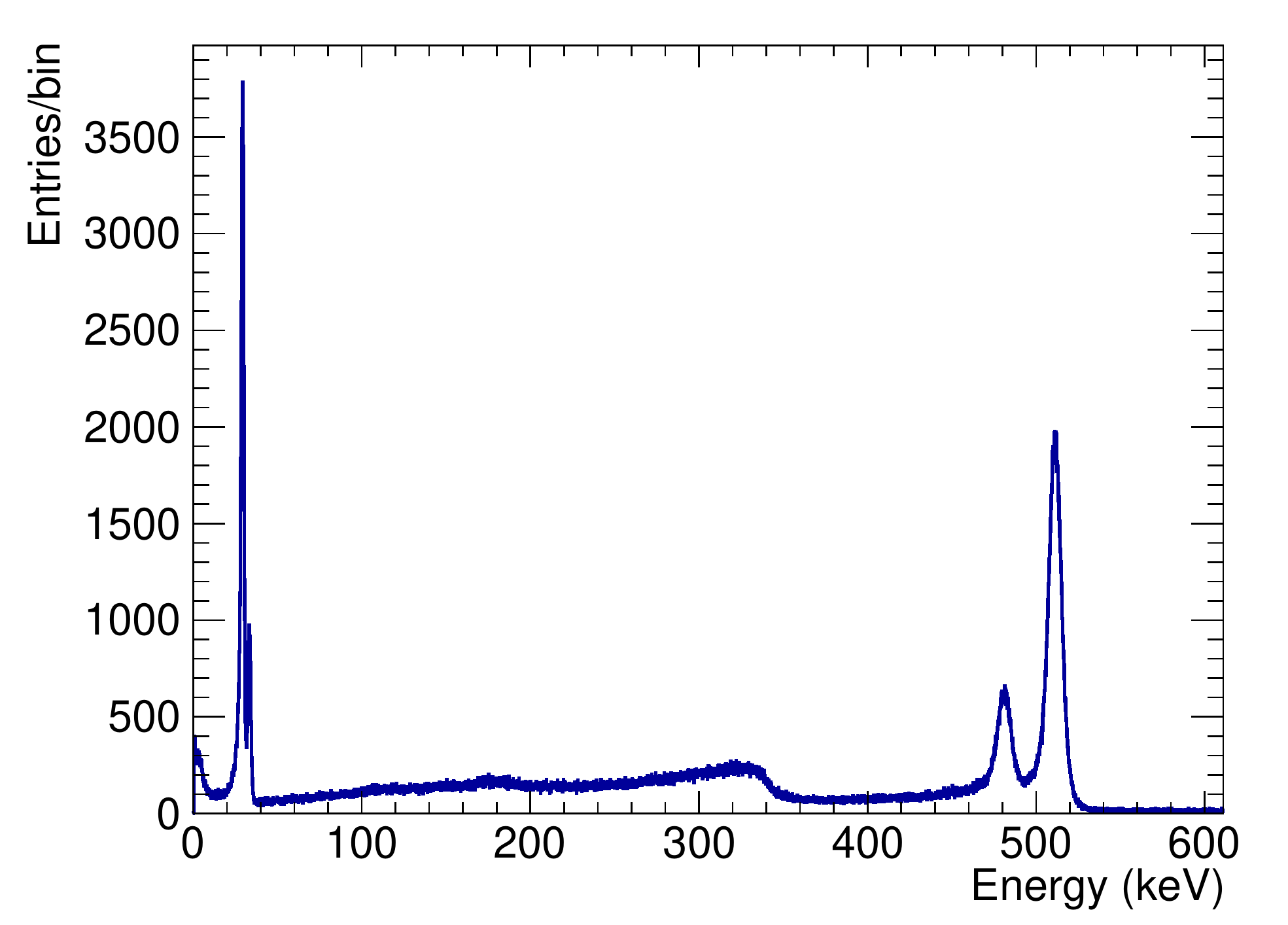}
\includegraphics[width=0.495\textwidth]{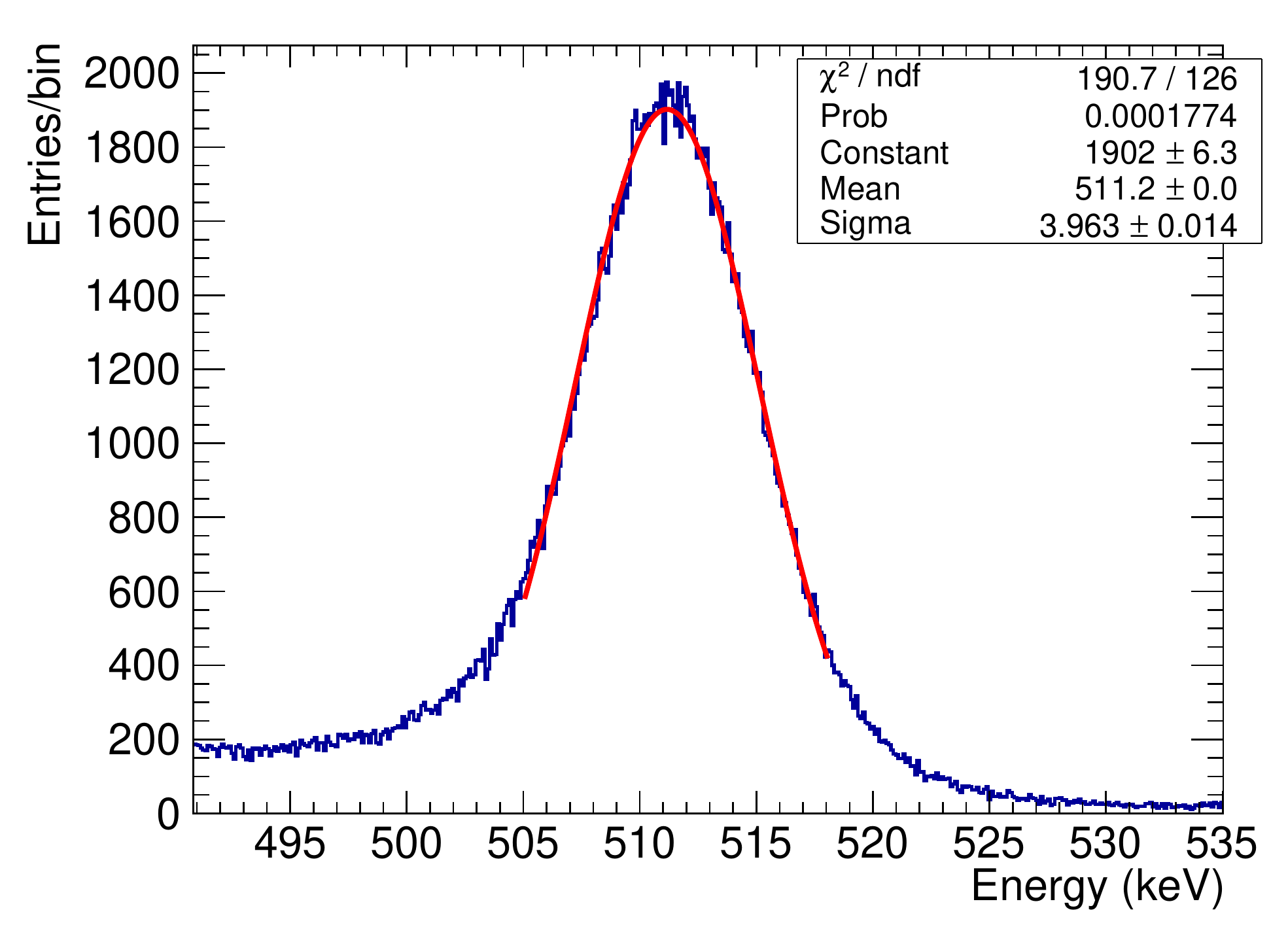}
\caption{Energy spectrum after spatial corrections (left) and detail
  of the photoelectric peak with fit values (right).}
\label{fig:spectrums}
\end{figure}

The geometrical dependence of the observed event charge has a number
of physical origins. Among these are: losses along the drift due to
electron attachment; losses at the edge of the fiducial region due to
inhomogeneities in the electric field; losses due to absorption of
light in the meshes and imperfect reflection and rotational symmetry
of the light tube. The first effects demand the continuous
recirculation and filtering of the gas to minimize attatchment while
the latter require the correction of the observed charge.

The dependence of the
observed energy on deposit position can be described by a geometric
factor parametrized as:
%%%
\begin{equation}
\eta(x,y,z) = Z(z') \cdot F(x,y) \label{eq:SpatialCorr}
\end{equation}
%%%
The $z$ dependence ($Z(z')$), due to attachment of the electrons in
the gas, is expected to be uncorrelated with other factors and can be treated independently. However, the dependence on
$x$ and $y$ position are sufficiently correlated that they should be
considered simultaneously.

\subsubsection{Z correction}
\label{subsub:zcorr}
The charge dependence on $z$ is well described by an exponential decay
scaling with the electron lifetime in the gas. The high quality of the
gas in these data runs results in a mean electron lifetime of greater
than 10~ms, implying a charge loss due to attachment of less than
3\%. Figure \ref{fig:zcorr} shows the charge dependence with z and how
the effect is corrected.

\subsubsection{XY correction}
\label{subsub:xycorr}
After the $z$ correction, correction for transverse inhomogeneities in the detector response was
performed using a 2 dimensional profile of the reconstructed charge in
the photoelectric peak as a function of reconstructed $xy$
position. Correction factors were computed for all positions using
Delaunay interpolation among separate bins (bins correspond
to a region with surface area of approximately $1.6\times 1.6$~cm$^2$)
and normalizing to the value for the bin at $R=0$. This
method was applied iteratively until convergence which was
generally achieved using 4 iterations.
Figure~\ref{fig:xyprofiles} shows the profile
after $z$ without $xy$ correction, the profile after the first iteration
and after asymptotic stability is reached.

After the corrections described, the resolution 
reaches 1.82\% FWHM in the photoelectric peak at 511~keV (figure \ref{fig:spectrums}).
This extrapolates to an energy resolution of 0.83\% FWHM at \Qbb .

\section{Summary and prospects}
\label{sec:conc}
NEXT-DEMO was designed as a proof of concept and test-bed of the NEXT-100 detector. In its current configuration, NEXT-DEMO fully implements the technologies envisaged for the experiment and, as such, represents a scale model of NEXT-100.
In the studies presented in this paper, the new tracking plane was used to improve the results obtained in
\cite{Alvarez:2012xda} significantly increasing the size of the fiducial region and understanding of event topologies. 

The detector has been operated over long periods of time and demonstrated stability and high gas quality; electron lifetimes in excess of 10~ms have been measured. At this level of attachment signals drifting over the 1~m drift distances (equivalent to $\sim$1~ms drift time) of NEXT-100 will suffer very little degradation. Additionally, the gains of TPB coating have been demonstrated and the stability of response validated. The SiPM array has been used to reconstruct event topologies and to demonstrate its power to improve energy resolution as well as reject background. The response of the photodetectors has been monitored and shown to be stable over the course of data taking.

The resolution achieved in the whole fiducial volume (an area representing almost 40\% of the fully instrumented volume) for the photoelectric interactions of gammas from a \NA\ source (511~keV) is 1.82\% FWHM. This energy resolution scales to a predicted FWHM resolution at \Qbb\ of 0.83\%, considerably better than NEXT target value.

Further work in preparation includes: the study of the energy scale using different energy sources; the study of the factors affecting the resolution and the development of algorithms for automatic characterisation of electron tracks.

%%%%%%%%%%%%%%%%%%%%%%%%%%%%%%%%%%%%%%%%%%%%%%%%%%%%%%%%%%%%
\acknowledgments
This work was supported by the following agencies and institutions:
the Ministerio de Econom\'ia y Competitividad of Spain under grants
CONSOLIDER-Ingenio 2010 CSD2008-0037 (CUP), FPA2009-13697-C04-04 and FIS2012-37947-C04;
the Director, Office of Science, Office of Basic Energy Sciences, of
the US Department of Energy under contract no.\ DE-AC02-05CH11231; and
the Portuguese FCT and FEDER through the program COMPETE, projects
PTDC/FIS/103860/2008 and PTDC/FIS/112272/2009. J.~Renner (LBNL) acknowledges the support of a
US DOE NNSA Stewardship Science Graduate Fellowship under contract
no.\ DE-FC52-08NA28752.

%%%%%%%%%%%%%%%%%%%%%%%%%%%%%%%%%%%%%%%%%%%%%%%%%%%%%%%%%%%%
\bibliographystyle{JHEP}
\bibliography{references}

\providecommand{\href}[2]{#2}\begingroup\raggedright\begin{thebibliography}{10}

\bibitem{Alvarez:2012haa}
{\bf NEXT} Collaboration, V.~\'Alvarez et~al., {\it {NEXT-100 Technical Design
  Report (TDR): Executive Summary}},  {\em JINST} {\bf 7} (2012) T06001,
  [\href{http://xxx.lanl.gov/abs/1202.0721}{{\tt arXiv:1202.0721}}].

\bibitem{GomezCadenas:2011it}
J.~J. G\'omez-Cadenas, J.~Mart\'in-Albo, M.~Mezzetto, F.~Monrabal, and
  M.~Sorel, {\it {The search for neutrinoless double beta decay}},  {\em Riv.\
  Nuovo Cim.} {\bf 35} (2012) 29--98,
  [\href{http://xxx.lanl.gov/abs/1109.5515}{{\tt arXiv:1109.5515}}].

\bibitem{GomezCadenas:2012jv}
J.~J. G\'omez-Cadenas, J.~Mart\'in-Albo, and F.~Monrabal, {\it {NEXT,
  high-pressure xenon gas experiments for ultimate sensitivity to Majorana
  neutrinos}},  {\em JINST} {\bf 7} (2012) C11007,
  [\href{http://xxx.lanl.gov/abs/1210.0341}{{\tt arXiv:1210.0341}}].

\bibitem{Nygren:2009zz}
D.~Nygren, {\it {High-pressure xenon gas electroluminescent TPC for
  $0\nu\beta\beta$ decay search}},  {\em Nucl.\ Instrum.\ Meth.\ A} {\bf 603}
  (2009) 337--348.

\bibitem{Alvarez:2011my}
{\bf NEXT} Collaboration, V.~Alvarez et~al., {\it {The NEXT-100 experiment for
  neutrinoless double beta decay searches (Conceptual Design Report)}},
  \href{http://xxx.lanl.gov/abs/1106.3630}{{\tt arXiv:1106.3630}}.

\bibitem{Alvarez:2012xda}
{\bf NEXT} Collaboration, V.~Alvarez et~al., {\it {Initial results of
  NEXT-DEMO, a large-scale prototype of the NEXT-100 experiment}},  {\em JINST}
  {\bf 8} (2013) P04002, [\href{http://xxx.lanl.gov/abs/1211.4838}{{\tt
  arXiv:1211.4838}}].

\bibitem{Alvarez:2012hu}
{\bf NEXT} Collaboration, V.~\'Alvarez et~al., {\it {Ionization and
  scintillation response of high-pressure xenon gas to alpha particles}},  {\em
  JINST} {\bf 8} (2013) P05025, [\href{http://xxx.lanl.gov/abs/1211.4508}{{\tt
  arXiv:1211.4508}}].

\bibitem{Toledo:2011zz}
J.~Toledo, H.~Muller, R.~Esteve, J.~Monzo, A.~Tarazona, et~al., {\it {The
  Front-End Concentrator card for the RD51 Scalable Readout System}},  {\em
  JINST} {\bf 6} (2011) C11028.

\bibitem{Alvarez:2012ub}
V.~Alvarez, J.~Agramunt, M.~Ball, M.~Batalle, J.~Bayarri, et~al., {\it {SiPMs
  coated with TPB : coating protocol and characterization for NEXT}},  {\em
  JINST} {\bf 7} (2012) P02010, [\href{http://xxx.lanl.gov/abs/1201.2018}{{\tt
  arXiv:1201.2018}}].

\bibitem{5873750}
N.~Dinu, C.~Bazin, V.~Chaumat, C.~Cheikali, A.~Para, V.~Puill, C.~Sylvia, and
  J.~F. Vagnucci, {\it {Temperature and bias voltage dependence of the MPPC
  detectors}},  in {\em Nuclear Science Symposium Conference Record (NSS/MIC),
  2010 IEEE}, pp.~215--219, 2010.

\bibitem{Rodriguez:2012IEEE}
{J. Rodr\'iguez, M. Querol, J. D\'iaz, J.J. G\'omez-Cadenas, D. Lorca, V.
  \'Alvarez, A. Gil}, {\it {Mass production automated test system for the NEXT
  SiPM tracking plane}},  in {\em {2012 IEEE NSS-MI Conference Record}},
  {2012}.

\bibitem{Alvarez:2012xdb}
{\bf NEXT} Collaboration, V.~Alvarez et~al., {\it {Design and characterization
  of the SiPM tracking system of NEXT-DEMO, a demonstrator prototype of the
  NEXT-100 experiment}},  {\em JINST} {\bf 8} (2013) T05002,
  [\href{http://xxx.lanl.gov/abs/1206.6199}{{\tt arXiv:1206.6199}}].

\bibitem{Janesick:2001}
J.~Janesick, {\em {Scientific Charge Coupled Devices}}.
\newblock {SPIE Press}, 2001.

\bibitem{NISTESTAR}
M.~Berger, J.~Coursey, M.~Zucker, , and J.~Chang, ``{(2005) ESTAR, PSTAR, and
  ASTAR: Computer Programs for Calculating Stopping-Power and Range Tables for
  Electrons, Protons, and Helium Ions (version 1.2.3)}.'' {Originally published
  as: Berger, M.J., NISTIR 4999, National Institute of Standards and
  Technology, Gaithersburg, MD (1993).}, February, 2013.
\newblock {URL: http://physics.nist.gov/Star}.

\end{thebibliography}\endgroup

\end{document}